\documentstyle[psfig]{mn}

\newif\ifAMStwofonts
\AMStwofontstrue

\newcommand{\be}{\begin{equation}}
\newcommand{\ee}{\end{equation}}
\newcommand{\ba}{\begin{eqnarray}}
\newcommand{\ea}{\end{eqnarray}}
\newcommand{\brr}{\begin{array}}
\newcommand{\err}{\end{array}}
\newcommand{\bc}{\begin{center}}
\newcommand{\ec}{\end{center}}
\newcommand{\hm}{\,h^{-1}{\rm Mpc}}

\newcommand{\mincir}{\raise
  -2.truept\hbox{\rlap{\hbox{$\sim$}}\raise5.truept \hbox{$<$}\ }}
\newcommand{\magcir}{\raise
  -2.truept\hbox{\rlap{\hbox{$\sim$}}\raise5.truept \hbox{$>$}\ }}
\newcommand{\siml}{\raise
  -2.truept\hbox{\rlap{\hbox{$\sim$}}\raise5.truept \hbox{$<$}\ }}
\newcommand{\simg}{\raise
  -2.truept\hbox{\rlap{\hbox{$\sim$}}\raise5.truept \hbox{$>$}\ }}

\ifoldfss
\ifCUPmtlplainloaded \else
\NewTextAlphabet{textbfit} {cmbxti10} {}
\NewTextAlphabet{textbfss} {cmssbx10} {}
\NewMathAlphabet{mathbfit} {cmbxti10} {} 
\NewMathAlphabet{mathbfss} {cmssbx10} {} 
\fi
\ifAMStwofonts
\ifCUPmtlplainloaded \else
\NewSymbolFont{upmath} {eurm10}
\NewSymbolFont{AMSa} {msam10}
\NewMathSymbol{\upi}     {0}{upmath}{19}
\NewMathSymbol{\umu}     {0}{upmath}{16}
\NewMathSymbol{\upartial}{0}{upmath}{40}
\NewMathSymbol{\leqslant}{3}{AMSa}{36}
\NewMathSymbol{\geqslant}{3}{AMSa}{3E}

\fi
\fi
\fi 

\ifnfssone
\newmathalphabet{\mathit}
\addtoversion{normal}{\mathit}{cmr}{m}{it}
\addtoversion{bold}{\mathit}{cmr}{bx}{it}
\newmathalphabet{\mathbfit} 
\addtoversion{normal}{\mathbfit}{cmr}{bx}{it}
\addtoversion{bold}{\mathbfit}{cmr}{bx}{it}
\newmathalphabet{\mathbfss} 
\addtoversion{normal}{\mathbfss}{cmss}{bx}{n}
\addtoversion{bold}{\mathbfss}{cmss}{bx}{n}
\ifAMStwofonts
\ifCUPmtlplainloaded \else
\UseAMStwoboldmath
\makeatletter
\new@mathgroup\upmath@group
\define@mathgroup\mv@normal\upmath@group{eur}{m}{n}
\define@mathgroup\mv@bold\upmath@group{eur}{b}{n}
\edef\UPM{\hexnumber\upmath@group}
\new@mathgroup\amsa@group
\define@mathgroup\mv@normal\amsa@group{msa}{m}{n}
\define@mathgroup\mv@bold\amsa@group{msa}{m}{n}
\edef\AMSa{\hexnumber\amsa@group}
\makeatother
\mathchardef\upi="0\UPM19
\mathchardef\umu="0\UPM16
\mathchardef\upartial="0\UPM40
\mathchardef\leqslant="3\AMSa36
\mathchardef\geqslant="3\AMSa3E
\fi
\fi
\fi 

\ifnfsstwo
\DeclareMathAlphabet{\mathbfit}{OT1}{cmr}{bx}{it}
\SetMathAlphabet\mathbfit{bold}{OT1}{cmr}{bx}{it}
\DeclareMathAlphabet{\mathbfss}{OT1}{cmss}{bx}{n}
\SetMathAlphabet\mathbfss{bold}{OT1}{cmss}{bx}{n}
\ifAMStwofonts
\ifCUPmtlplainloaded \else
\DeclareSymbolFont{UPM}{U}{eur}{m}{n}
\SetSymbolFont{UPM}{bold}{U}{eur}{b}{n}
\DeclareSymbolFont{AMSa}{U}{msa}{m}{n}
\DeclareMathSymbol{\upi}{0}{UPM}{"19}
\DeclareMathSymbol{\umu}{0}{UPM}{"16}
\DeclareMathSymbol{\upartial}{0}{UPM}{"40}
\DeclareMathSymbol{\leqslant}{3}{AMSa}{"36}
\DeclareMathSymbol{\geqslant}{3}{AMSa}{"3E}
\fi
\fi
\fi 

\ifCUPmtlplainloaded \else
\ifAMStwofonts \else 
\def\upi{\pi}
\def\umu{\mu}
\def\upartial{\partial}
\fi
\fi

\title[Cooling and heating the ICM] {Cooling and heating the ICM in hydrodynamical simulations} 
\author[L. Tornatore et al.]{
L. Tornatore$^1$,
S. Borgani$^1$, 
V. Springel$^2$,
F. Matteucci$^1$,
N. Menci$^3$, 
G. Murante$^4$
\\~\\
$^1$ Dipartimento di Astronomia
dell'Universit\`a di Trieste, via Tiepolo 11, I-34131 Trieste, Italy
(tornatore, borgani, matteucci @ts.astro.it)\\ 
$^2$ Max-Planck-Institut f\"ur Astrophysik, Karl-Schwarzschild Strasse 1,
Garching bei M\"unchen, Germany (volker@mpa-garching.mpg.de)\\
$^3$ INAF, Osservatorio Astronomico di Roma, via dell'Osservatorio, I-00040
Monteporzio, Italy (menci@coma.mporzio.astro.it) \\ 
$^4$ INAF, Osservatorio Astronomico di Torino, Strada Osservatorio 20,
Pino Torinese, I-10025 Italy (giuseppe@to.astro.it)\\
}
\pubyear{2002}

\begin{document}
\label{firstpage}
\maketitle

\begin{abstract}
We discuss Tree+SPH simulations of galaxy clusters and groups, aimed
at studying the effect of cooling and non--gravitational heating on
observable properties of the intra--cluster medium (ICM). We simulate
at high resolution four group- and cluster-sized halos, with virial
masses in 
the range (0.2--4)$\times 10^{14}M_\odot$, extracted from a
cosmological simulation of a flat $\Lambda$CDM model. We discuss the
effects of using different SPH implementations and show that high
resolution is mandatory to correctly follow the cooling pattern of the
ICM. Our recipes for non--gravitational heating release energy to the
gas either in an impulsive way, at some heating redshift, or by
modulating the heating as a function of redshift according to the star
formation history predicted by a semi--analytic model of galaxy
formation. Our simulations demonstrate that cooling and
non--gravitational heating exhibit a rather complex interplay in
determining the properties of the ICM: results on the amount of star
formation and on the $X$--ray properties are sensitive not only to the
amount of heating energy, but also depend on the redshift at which it
is assigned to gas particles. All of our heating schemes which
correctly reproduce the $X$--ray scaling properties of clusters and
groups do not succeed in reducing the fraction of collapsed gas below
a level of 20 (30) per cent at the cluster (group) scale, which
appears to be in excess of observational constraints. Finally, gas
compression in cooling cluster regions causes an increase of the
temperature and a steepening of the temperature profiles, independent
of the presence of non-gravitational heating processes. This is
inconsistent with recent observational evidence for a decrease of gas
temperature towards the center of relaxed clusters. Provided these
discrepancies persist even for a more refined modeling of energy
feedback from supernova or AGN, they may indicate that some basic physical
process is still missing in hydrodynamical simulations.
\end{abstract}


\section{Introduction}
The simplest picture to describe the thermal properties of the
intra--cluster medium (ICM) is based on the assumption that gas
heating occurs only by the action of gravitational processes, such as
adiabatic compression from gravitational collapse, and by
hydrodynamical shocks from supersonic accretion (Kaiser 1986). Since
gravity does not have characteristic scales, this model predicts
that galaxy systems of different mass look like scaled versions of each
other.
Under the assumptions of thermal bremsstrahlung emissivity and
hydrostatic equilibrium, this model provides precise predictions for
$X$--ray scaling properties of galaxy systems: {\bf (a)} $L_X\propto
T^2(1+z)^{3/2}$ for the shape and evolution of the relation between
$X$--ray luminosity and gas temperature; {\bf (b)} $S\propto
T(1+z)^{-2}$ for the entropy-temperature relation, where
$S=T/n_e^{2/3}$ is the gas entropy and $n_e$ is electron number
density; {\bf (c)} $M\propto T^{3/2}$ for the relation between total
cluster virial mass and temperature, with normalization determined by
the parameter $\beta=\mu m_p \sigma_v^2/k_BT$ ($\mu=0.59$ mean
molecular weight for primordial composition; $m_p$: proton mass; and
$\sigma_v$: line-of-sight velocity dispersion). Numerical simulations
that only include gravitational heating showed that $\beta\simeq
1$--1.3 (e.g. Navarro, Frenk \& White 1995; Evrard, Metzler \& Navarro
1996; Bryan \& Norman 1998; Eke et al. 1998b; Borgani, Governato,
Wadsley et al. 2002, BGW hereafter).

A number of observational facts demonstrate that this picture is too
simplistic, thus calling for the consideration of extra physics in the
description of the ICM. The $L_X$--$T$ relation is found to be steeper
than predicted, with $L_X\propto T^{\sim 3}$ at $T_X> 2$ keV (e.g.,
White, Jones \& Forman 1997; Markevitch 1998; Arnaud \& Evrard 1999;
Ettori, De Grandi \& Molendi 2002), possibly approaching the
self--similar scaling only for the hottest systems with $T\magcir 8$
keV (Allen \& Fabian 1998).  Evidences also emerged for this relation
to further steepen for colder groups, $T\mincir 1$ keV (e.g., Ponman
et al. 1996; Helsdon \& Ponman 2000; Mulchaey 2000). Furthermore, no
evidence for a strong positive evolution of the $L_X$--$T$ relation
has been found to date out to $z\sim 1$ (e.g., Mushotzky \& Scharf
1997; Reichart et al. 1999; Fairley et al. 2000; Borgani et al. 2001a;
Holden et al. 2002; Novicki, Sornig \& Henry 2002; cf. also Vikhlinin
et al. 2002). As for the $S$--$T$ relation, Ponman, Cannon \& Navarro
(1999) found from ROSAT and ASCA data an excess of entropy within the
central regions of $T\mincir 2$ keV systems (see also Lloyd--Davis et
al. 2000, Finoguenov et al. 2002a), possibly approaching the value
$S\sim 100$ keV cm$^2$ for the coldest groups. Finally, a series of
evidences, based on ASCA (e.g., Horner, Mushotzky \& Scharf 1999;
Nevalainen, Markevitch \& Forman 2000; Finoguenov, Reiprich \&
B\"ohringer 2001b), Beppo--SAX (Ettori et al. 2002) and Chandra (Allen,
Schmidt \& Fabian 2001) data, shows that the observed $M$--$T$
relation has a $\sim 40$ per cent lower normalization than predicted
by simulations that only include gravitational heating.

In the attempt of interpreting these data, theoreticians are currently
following two alternative routes, based either on introducing
non--gravitational heating of the ICM or on alluding to the effects of
radiative cooling.

An episode of non--gravitational heating, occurring before or during
the gravitational collapse, has the effect of increasing the entropy
of the gas, preventing it from reaching high densities in the central
cluster regions and suppressing its $X$--ray emissivity (e.g., Evrard
\& Henry 1991, Kaiser 1991; Bower 1997). For a fixed amount of
specific heating, the effect is larger for poorer systems, i.e. when
the extra energy per gas particle is comparable to the halo virial
temperature. This produces both an excess entropy and a steeper
$L_X$--$T$ relation (e.g., Cavaliere, Menci \& Tozzi 1998; Balogh,
Babul \& Patton 1999; Tozzi \& Norman 2001). Arguments based on
semi--analytical work (e.g., Tozzi \& Norman 2001) and numerical
simulations (Bialek, Evrard \& Mohr 2001; Brighenti \& Mathews 2001;
Borgani et al. 2001b, 2002) suggest that a specific heating energy of
$E_h\sim 1$ keV/part or, equivalently, a pre--collapse entropy floor
of $S\sim 100$ keV cm$^2$, can account for the observed $X$--ray
properties of galaxy systems (cf. also Babul et al. 2002, Finoguenov
et al. 2002a for arguments suggesting a stronger pre--heating). Yet,
the origin for this energy has still to be determined. Energy release
from supernovae feedback has been advocated as a possibility (e.g.,
Bower et al. 2001; Menci \& Cavaliere 2001). Using the abundance of
heavy elements of the ICM as a diagnostic for the past history of the
star formation within clusters (e.g., Renzini 1997; Kravtsov \& Yepes
2000; Pipino et al. 2002; Valdarnini 2002), a number of studies
concluded that SN may fall short in providing the required
extra--energy budget (cf. also Finoguenov, Arnaud \& David 2001a). The
other obvious candidate is represented by energy from AGN (e.g.,
Valageas \& Silk 1999; Wu, Fabian \& Nulsen 2000; Mc Namara et
al. 2000; Nath \& Roychowdhury 2002; Cavaliere, Lapi \& Menci
2002). In this case, the large amount of energy that is available
requires some degree of tuning of the mechanisms responsible for its
conversion into thermal energy of the gas.  While a suitable amount of
non--gravitational heating can account for the observed $L_X$--$T$
relation and entropy excess, the $M$--$T$ relation is only marginally
affected by extra heating (e.g. BGW), thus leaving the discrepancy
between observed and predicted relation unresolved.

As for cooling, its effect is to selectively remove those low--entropy
particles from the diffuse $X$--ray emitting phase which have cooling
times shorter than the Hubble time (e.g., Voit \& Bryan 2002; Wu \&
Xue 2002). Conversion of cooled gas into collisionless stars decreases
the central gas density and, at the same time, the resulting lack of
pressure support causes higher--entropy shocked gas to flow in from
the outskirts of the cluster or group. As a result, the $X$--ray
luminosity is suppressed, while the entropy increases, much like in a
pre--heating scenario (Pearce et al. 2001; Muanwong et al. 2002;
Dav\'e, Katz \& Weinberg 2002). However, by its nature, cooling is
known to be a runaway process: cooling causes gas to be accumulated
into dense structures, and the efficiency of cooling increases with
gas density. As a result, most simulations consistently predict a
significant fraction of gas to be converted into cold ``stars'',
$f_{\rm cold}\magcir 30$ per cent (e.g., Suginohara \& Ostriker 1998;
Lewis et al. 2000; Yoshida et al. 2002; BGW), while observations
indicate a considerably lower value of $f_{\rm cold}\mincir 10$ per
cent (e.g., Balogh et al. 2001; Wu \& Xue 2002).

This suggests that in real clusters some source of extra heating is
increasing the entropy of the gas, preventing overcooling. Voit et al
(2002) have developed a semi--analytical approach to derive $X$--ray
observable properties of the ICM in the presence of both cooling and
extra heating. Based on this approach, these authors found that
cooling and a modest amount of extra heating are able to account for
basically all the $X$--ray ICM observables. Oh \& Benson (2002)
pointed out that pre--heating is needed to increase the cooling time
and prevent overcooling, by suppressing the gas supply to galaxies
(see also Finoguenov et al. 2002b). It is however clear that,
as for any analytical approach, suitable assumptions and
approximations are needed to choose criteria for removing cooled
gas from the hot diffuse phase, and to follow the complex dynamics of
cooling/heating of gas during the process of cluster formation.

Muanwong et al. (2002) and Kay, Thomas \& Theuns (2002) used
hydrodynamical simulations within a cosmological box to study the
interplay of gas cooling and a few prescriptions for
non--gravitational heating. As a general result, they found that
increasing the heating can suppress the amount of cooled gas.  While
the choice of simulating a whole cosmological box has the advantage of
providing a large statistics of groups and clusters, it also severely
limits the available mass and force resolution. On the other hand, by
the very nature of cooling, increasing the mass resolution allows to
follow the formation of smaller halos at progressively larger
redshift, where cooling and, potentially, star formation are
particularly efficient. As a consequence, unless very high mass
resolution is achieved, cooling in simulations can be significantly
underestimated (e.g., Balogh et al. 2001).

In this paper, we follow the alternative approach of simulating at
very high resolution a limited number of group-- and
cluster--sized halos selected from a cosmological box, and
we widen the explored range of possible patterns for
non--gravitational heating (see also BGW). While this limits our
ability to precisely calibrate shape and scatter of $X$--ray scaling
relations, we are able to increase the resolution in the most
interesting regions of the gas distribution. Indeed, the simulations
presented in this paper are among the highest resolution attempts
realized so far to follow the structure of gas cooling within groups
and clusters in the presence of a variety of schemes for extra gas
heating.
Furthermore, we also investigate how the cooling efficiency depends
both on numerical resolution and on details of the SPH implementation.

The structure of this paper is as follows. After providing a short
description of the code, we present in Section 2 the procedure to
simulate individual halos at high resolution and discuss the main
characteristics of the four selected halos. In Section 3, we discuss
the results on the cold fraction. Here we will concentrate on showing
how this fraction depends on numerical resolution, integration scheme
and removal of cold dense particles from the SPH computation (star
formation). Finally, we present the adopted schemes for
non--gravitational gas heating and discuss their impact on the
resulting cold fraction and pattern of star formation.  In Section 4,
we present the predictions on $X$--ray properties of clusters and
groups from our simulations, namely the entropy--temperature, the
luminosity--temperature and the mass--temperature relations. Finally,
we discuss our main results and draw conclusions in Section 5.

\section{The simulations}
\subsection{The code}
Our simulations are realized with {\small GADGET}\footnote{\tt
http://www.mpa-garching.mpg.de/gadget}, a parallel tree N--body/SPH
code (Springel, Yoshida \& White 2001), with fully adaptive time--step
integration. Gas cooling in the SPH part of the code is implemented
following Katz, Weinberg \& Hernquist (1996, KWH
hereafter). Specifically, the abundances of ionic species are computed
by assuming collisional equilibrium for a gas of primordial
composition (mass--fraction $X=0.76$ of hydrogen and $1-X=0.24$ of
helium). Since we not follow metal production from star--formation, we
do not include the effect of metals on the cooling function. We
include the effect of a time--dependent uniform UV background (e.g.,
Haardt \& Madau 1999), although its effect is only very small for the
massive objects we focus on in this study. We set the number of
neighbors for SPH computations to 32, allowing the SPH smoothing
length to drop at most to the value of the gravitational softening
length of the gas particles.

\subsection{The simulated structures}
We simulate four halos at high resolution, which are extracted from a
low--resolution DM only simulation within a box of $70\hm$ on a side,
for a cosmological model with $\Omega_m=0.3$, $\Omega_\Lambda=0.7$,
Hubble constant $H_0=70$ km s$^{-1}$ Mpc$^{-1}$ and normalization
$\sigma_8=0.8$, consistent with recent determinations of the number
density of nearby clusters (Pierpaoli et al. 2002, and references
therein). As for the baryon content, we assume $\Omega_{\rm
bar}=0.019\,h^{-2}$ (e.g., Burles \& Tytler 1998). This choice of
$\Omega_{\rm bar}$ corresponds to $f_{\rm bar}\simeq 0.13$ for the
cosmic baryon fraction, which, for the assumed cosmology, is
consistent with the value measured from cluster observations (e.g.,
Ettori 2002, and references therein).

The most massive halo we selected corresponds to a Virgo--like
cluster, with virial mass of about $4\times 10^{14}M_\odot$ (as usual,
we call ``virial'' the mass within the radius encompassing the virial
overdensity computed for the simulated cosmology; e.g. Eke et
al. 1998a). This turns out to be the most massive system extracted
from the simulation box. In the following, we will refer to this
system as the ``Virgo'' cluster. The other three halos, which have
been extracted from
a single
Lagrangian region, correspond to groups in the mass range
(2--6)$\times 10^{13}M_\odot$. In the following, we will refer to
these three structures as ``Group-1'', ``Group-2'' and ``Group-3''. We
provide in Table \ref{t:simul} the main characteristics of the
simulated structures.

We follow the technique originally presented by Katz \& White (1993)
to increase the mass resolution and to add short wavelength modes
within Lagrangian regions that contain the structures of interest. In
these high--resolution regions, particles are split into a dark matter
and a gaseous part, with mass ratio reflecting the value of the cosmic
baryon fraction.  Force and mass resolution are then gradually
degraded in the outer regions, so as to limit the computational cost,
while providing a correct representation of the large--scale tidal
field. The size of the regions selected at $z=0$, to be resimulated at
high resolution, typically corresponds to 10-20 Mpc in Lagrangian
space, and is always chosen to be large enough that no low-resolution
heavy particles contaminate the virial region of the simulated halos.

In order to assess numerical effects, structures have been simulated
at different mass and force resolutions. We fix three different mass
resolutions, which correspond to $m_{\rm gas}\simeq 2.5\times
10^9M_\odot$, $3.2\times 10^8M_\odot$ and $3.9\times 10^7M_\odot$ for
the mass of the gas particles, respectively. In the following, the
group runs with the smallest (intermediate) value of $m_{\rm gas}$,
and the Virgo runs with the intermediate (largest) $m_{\rm gas}$ will
be indicated as high--resolution (low--resolution) runs and labeled
with HR (LR). We do not discuss Virgo runs with the smallest $m_{\rm gas}$
and Group runs with the largest $m_{\rm gas}$ among this list of three
mass resolutions.  With these choices for the mass resolution, the HR
runs resolve the virial regions of the simulated structures with a
number of gas particles ranging from about 70,000 to about 185,000
(see Table \ref{t:simul}).  The redshift $z_i$ at which initial
conditions are generated is chosen such that the r.m.s. fluctuation in
the density field of the high--resolution region is $\sigma=0.1$ (on
the scale of the smallest resolved masses). With this requirement, we
have $z_i\simeq 65$.  As for the choice of the softening scale for the
computation of the gravitational force, we assume it to have a
constant value in comoving units down to $z=2$, and a constant value
in physical units at later epochs. The corresponding values of the
Plummer--equivalent softening scale at $z=0$ are $\epsilon_{\rm Pl}=10$, 5
and 2.5 kpc for the three different choices of $m_{\rm gas}$. This
choice has been dictated by the requirement of resolving halos down to
scales of about one percent of their virial radii, so as to
correctly follow the gas clumpiness and, therefore, to have convergent
estimates of the X--ray luminosity (e.g., Borgani et al. 2002).

\begin{figure*}
\centerline{\vbox{
\hbox{
\psfig{file=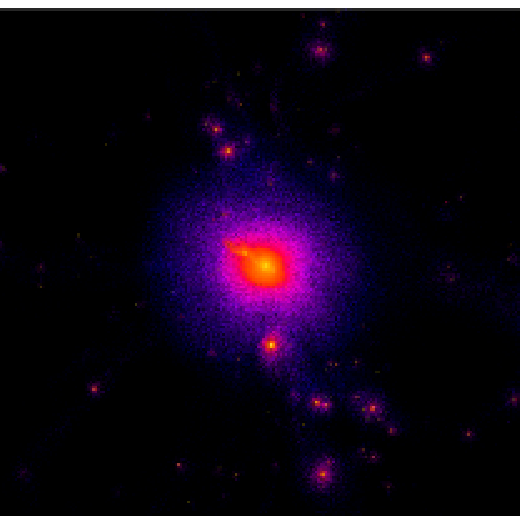,width=6.5cm} 
\psfig{file=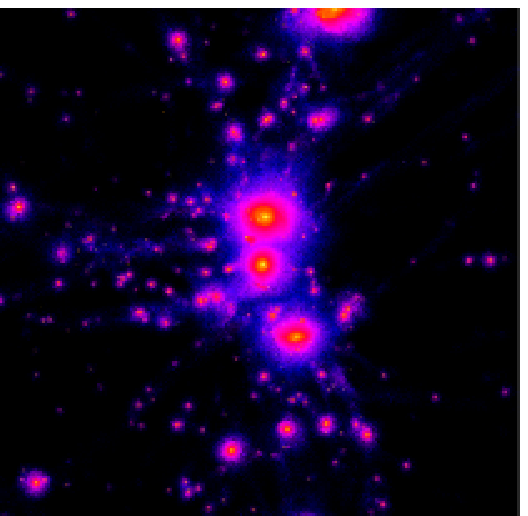,width=6.5cm}
}
\vspace{0.1cm}
\hbox{
\psfig{file=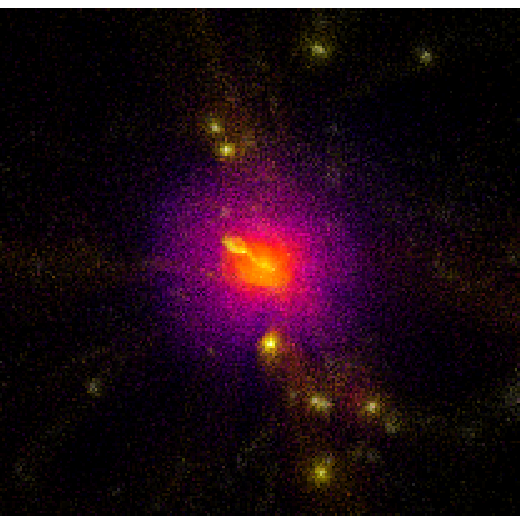,width=6.5cm} 
\psfig{file=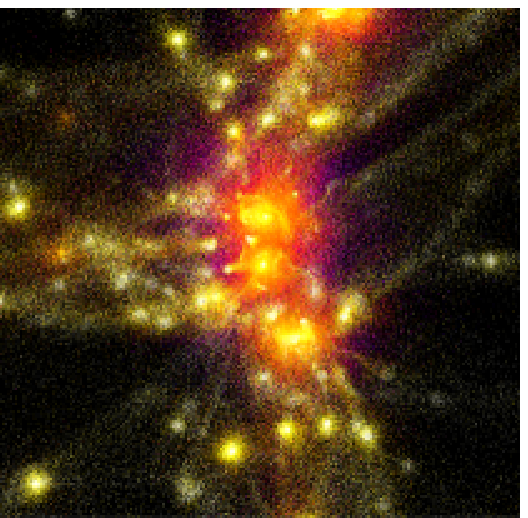,width=6.5cm}
}
}}
\caption{Maps of the gas density (upper panels) and of the gas entropy
(lower panels) for the Virgo run (left panels) and for the region
containing the groups (right panel), for the HR runs including cooling
and star formation (see text). The size of each
box is 10 Mpc, so as to show the environment of the simulated
systems. Brighter regions indicate higher gas density and lower
entropy in the upper and lower panels, respectively.}
\label{fi:maps} 
\end{figure*}

\begin{table}
\centering
\caption{Physical characteristics and numerical parameters of the
simulated halos in the HR runs. Column 2: total mass within the virial
radius at $z=0$ ($10^{13}M_\odot$); Column 3: virial radius (Mpc);
Column 4: total mass within $R_{500}$; Column 5: radius containing an
average density $\bar \rho=500 \rho_{\rm crit}$. Column 6: number of gas
particles within $R_{\rm vir}$; Column 7: Plummer-equivalent softening
parameter at $z=0$ ($h^{-1}$kpc).}
\begin{tabular}{lcccccc}
Run & $M_{\rm vir}$ & $R_{\rm vir}$ & $M_{500}$ & $R_{500}$ & $N_{\rm gas}$ & $\epsilon$\\
\hline 
Cluster & 39.4 & 1.90 & 23.3 & 0.94 & 1.5e5 & 5.0 \\ 
Group-1 & 5.98 & 1.01 & 3.43 & 0.49 & 1.8e5 & 2.5 \\ 
Group-2 & 2.52 & 0.76 & 1.60 & 0.38 & 7.8e4 & 2.5 \\ 
Group-3 & 2.35 & 0.74 & 1.35 & 0.36 & 7.1e4 & 2.5 \\
\hline
\end{tabular}
\label{t:simul}
\end{table}

We show in Figure~\ref{fi:maps} the gas density and entropy maps for
the HR runs of the cluster and of the group regions at $z=0$ for the
runs including cooling and star formation (see below). In the entropy
map of the Virgo cluster (lower left panel), we note a tail of low
entropy gas pointing toward the center. This feature is generated by a
merging sub-group, whose gas has been tidally stripped during the
first passage through the cluster virial region. The persistence of
low entropy for this gas indicates that it has been only recently
stripped and has still to thermalize within the cluster
environment. As apparent from the gas--density map (upper left panel),
this merging sub-halo gives rise only to a minor disturbance of the
gas density, thus marginally disturbing the relaxed dynamical status
of the cluster. As for the simulation of the region containing the
three groups, we note that they are placed along a filamentary
structure. Although they are still relatively isolated and separated
from each other by a few virial radii, their motion shows that they
are approaching each other and will merge to form a cluster--sized
structure in a few Gyrs. In general, these maps witness that a rich
variety of structures, emerging when high resolution is achieved, are
naturally expected to characterize the ICM, much like shown by high
resolution Chandra observations.

\section{Computing the collapsed gas fraction in cluster simulations}
\subsection{Introducing radiative cooling}
An important aspect when dealing with simulations that include cooling
concerns the detailed scheme of SPH implementation. Most standard
implementations integrate the specific thermal energy as an
independent variable, differing however in the detailed method used
for symmetrizing the pairwise hydrodynamic forces between gas particle
pairs, where either a simple arithmetic or a geometric mean form the
most common choices (e.g., Weinberg, Hernquist \& Katz 1997; Dav\`e et
al. 1999; White, Hernquist \& Springel 2001).  While these SPH
implementations conserve energy and momentum, Springel \& Hernquist
(2002, SH02 hereafter) have shown that several of the commonly used
SPH implementations are characterized by a spurious loss of specific
entropy in strongly cooling regions, an effect which can be
particularly severe at low resolution, and which is stronger when the
geometrical scheme for hydrodynamical force symmetrization is
adopted. This problem is essentially due to spurious coupling between
cool dense particles, which should have virtually left the collisional
phase, and neighboring hot gas particles, which still belong to the diffuse
phase. In order to avoid the resulting spurious overcooling, different
techniques have been suggested by several authors (e.g., Pearce et
al. 2001; Marri et al. 2002).

SH02 proposed a new SPH implementation based on integrating the
specific entropy as an independent thermodynamic variable, an approach
which explicitly conserves entropy in non--shock regions. Using a
variational principle to derive the SPH equations of motion, they also
showed that this new formulation removes any ambiguity in the choice
of symmetrization and conserves energy, even when adaptive smoothing
lengths are used.

Since one of the main purposes of this paper is to investigate the
properties of gas cooling in galaxy clusters, we will study below by
how much differences in the SPH implementation can change the
resulting fraction of cold gas.  Adopting the naming convention of
SH02, we refer to a standard SPH implementation with geometric
symmetrization as ``geometrical'', and to one with arithmetic
symmetrization as ``arithmetic'', while the new formulation of SH02
will be referred to as ``entropy--conserving''.

\begin{figure}
\centerline{
\psfig{file=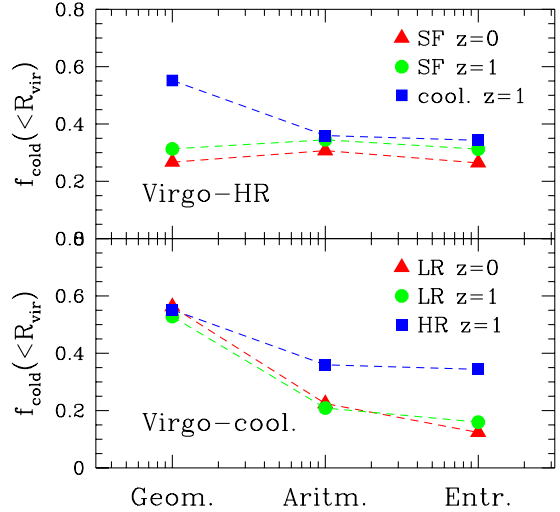,width=8cm} 
}
\vspace{-0.5truecm}
\caption{The fraction of collapsed gas within the virial radius for
the Virgo run, for the three different schemes of SPH
implementation. The lower panel refers to runs including cooling but
not star formation. The upper panel shows the effect of introducing
star formation for the HR runs.  }
\label{fi:fc_cool} 
\end{figure}

\subsection{Introducing star formation}
Star formation is introduced as an algorithm to remove dense cold gas
particles from the SPH computation, treating them as collisionless
``stars''. We follow the recipe originally proposed by KWH. According
to this recipe, a gas particle is eligible to form stars if the
following conditions are met: {\em (i)} locally convergent flow,
$\nabla \cdot {\bf v}<0$; {\em (ii)} Jeans unstable, i.e. locally
determined sound crossing time longer than dynamical crossing time;
{\em (iii)} gas overdensity exceeding a critical overdensity value,
$\delta_{g}>55$; {\em (iv)} local number density of hydrogen atoms
$n_{H}>0.1$ cm$^{-3}$.  

Once a particle is eligible to form stars, its star formation rate
(SFR) is given by ${\rm d} \ln \rho_g/{\rm d}t=-c_*/t_g$, where $t_g$
is the minimum between the local gas--dynamical time--scale,
$t_{\rm dyn}=(4\pi G\rho_g)^{-1/2}$, and the local cooling time--scale. We
assumed $c_*=0.1$ for the parameter regulating the rate of conversion
of cold gas into stars, and verified with a low--resolution simulation
of the Virgo cluster that basic results are left essentially unchanged
by taking instead $c_*=0.01$.  A gas particle eligible for star
formation is assumed to be gradually converted into a star particle,
according to the above SFR. Instead of creating a new star particle
for every star--formation (SF) instance, each gas particle undergoing
SF behaves in a ``schizophrenic'' way, with its stellar part feeling
only gravity (see, e.g., Mihos \& Hernquist 1994). Once the SPH mass
fraction decreases to 10 per cent, it is dissolved into SPH neighbors,
thus leaving a purely stellar particle.

We also follow the recipe by KWH to compute the energy feedback from
the SN associated with the star--formation produced in the
simulations. Assuming a Miller--Scalo (1979) initial mass function
(IMF), we compute the number of stars with mass $>8\,M_\odot$, which
we identify with instantaneously exploding SN. After assuming that
each SN releases $10^{51}$ ergs, the resulting amount of energy per
formed stellar mass turns out to be $7\times 10^{48}$ ergs
$M_\odot^{-1}$. While the approximation of instantaneous explosion can
be justified for type-II SN, due to the short life--time of their
progenitor stars, it is not valid for type-Ia SN, which have stellar
progenitors of smaller masses and much longer life--times (e.g. Lia,
Portinari \& Carraro 2001; Pipino et al. 2002; Valdarnini 2002). The
resulting feedback energy is assigned as thermal energy to the
star--forming gas particles. This scheme for SN feedback is known to
thermalize a negligible amount of energy in the diffuse medium, since
it acts mostly on cold dense particles which rapidly radiate away the
feedback energy as a consequence of their short cooling time. In the
following, we show results based on including this scheme for SN
feedback while bearing in mind that it causes only negligible
differences compared to simulations that lack any stellar feedback.
In Section 3.3 we shall discuss a different SN feedback scheme, based
on the predictions of semi--analytical modelling of galaxy formation.

\begin{figure}
\vspace{-3.5cm}
\centerline{
\psfig{file=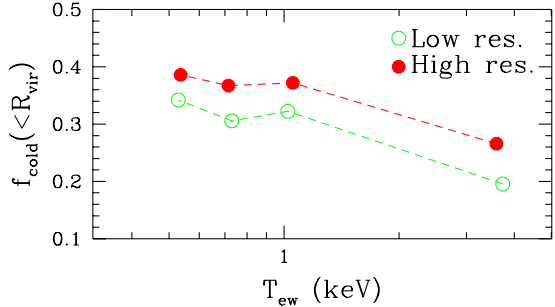,width=8cm} 
}
\vspace{-0.5truecm}
\caption{The fraction of cold gas as a function of the
emission--weighted temperature, $T_{\rm ew}$, of the simulated structures. 
Results at two different resolutions are shown for the simulations of
the groups.}
\label{fi:fc_star}
\end{figure}

The effect of including star formation on the fraction of collapsed
gas in the Virgo cluster run is shown in Figure
\ref{fi:fc_cool}. Besides the population of collisionless star
particles, we also define as belonging to the cold phase all the SPH
particles which have overdensity $\delta_{\rm gas}>500$ and
temperature $T< 3\times 10^4$ K (see also Croft et al. 2001, Borgani
et al. 2002).  At $z=1$, the star formation simulation produce $f_{\rm
cold}\simeq 25$--35 per cent of collapsed gas, with a weak dependence
on the integration scheme. However, in the cooling-only simulations,
where collapsed gas is not converted to stars, the ``geometric''
scheme leads to substantially larger values, indicating a numerical
overcooling problem in this method.  This effect is absent in the
entropy--conserving scheme, which proves effective in suppressing
spurious cooling in the absence of an explicit SF scheme. Note that
$f_{\rm cold}$ is seen to slightly decrease at later epochs, which is
as a consequence of a reduction of the rate at which cooling and star
formation proceeds with respect to diffuse gas accretion.

In Figure \ref{fi:fc_star} we show the trend of $f_{\rm cold}$ against
the emission--weighted temperature, which is defined here as 
\be
T_{ew}\,=\,{\sum_{i} \rho_i\,T_i^{3/2}\over \sum_{i}
\rho_i\,T_i^{1/2}}\,,
\label{eq:tew}
\ee 
where $\rho_i$ is the gas density carried by the $i$-th SPH
particle. This definition is valid only for bremsstrahlung emissivity,
although final values of $T_{ew}$ are left essentially unchanged if we
account for the contribution from metal lines.

For the groups, we plot results for
both the LR and the HR runs. The trend toward a higher $f_{\rm cold}$
in colder systems is a consequence of the shorter cooling times of the
associated DM halos, which makes cooling to proceed faster in
lower--mass systems. Quite apparently, increasing the resolution
causes a $\sim 20$ per cent or a $\sim 50$ per cent increase of
$f_{\rm cold}$ at the group and at the cluster scales,
respectively. The effect of numerical resolution is also shown in the
left panel of Figure \ref{fi:sfr}. In this figure we plot the density
of star--formation rate (SFR) within the virial region of the Virgo
cluster and of the Group-1. Once the same mass resolution is used for
the simulation of these two structures, the resulting SFRs are quite
similar. Increasing the resolution of the group simulation allows to
resolve smaller halos forming at higher $z$, where additional star
formation can take place. As a result, the peak in the SFR moves from
$z\simeq 2$ to $z\simeq 3$ and then declines more gently, while
recovering the same shape at lower redshift. Note that the integrated
star formation rate is dominated by the contribution from these low
redshifts, where most of the physical time is being spent.  The
resolution achieved in the HR runs is sufficient to resolve ``galaxy''
halos well below $L_*$. Therefore, we are confident that we are
obtaining nearly converged estimates of the collapsed gas fraction,
at least when the highest mass resolution is used. At the same time,
our results should be considered as a warning on the interpretation
of simulations that lack the resolution to adequately follow gas
cooling.

In summary, our simulations demonstrate that cold fractions as large
as $f_{\rm cold}=25$--35 per cent should be expected when radiative
cooling and star formation are considered. These values are larger
than the observed $\sim 10\%$ fraction of cold gas in clusters (e.g.,
Balogh et al. 2001). This calls for the need to introduce a
suitable scheme of non--gravitational energy injection, allowing a
regulation of the runaway cooling process.

\subsection{Introducing extra heating}
The SN feedback recipe that we discussed so far, where thermal energy
is deposited into cold gas, does not produce any sizable effect. In
order to overcome this problem, many different schemes have been
proposed. All these schemes attempt to prevent feedback energy from
being quickly radiated away, for example by suitably parameterizing
``sub--grid'' physics, such as the multi-phase structure of the
interstellar medium or galactic winds (e.g., Kay et al. 2000; Springel
\& Hernquist 2002; Marri \& White 2002).

Here we present different phenomenological approaches for
non--gravitational heating.  Rather than predicting feedback from the
star formation actually produced in the simulations, these schemes are
designed to shed light on how much extra energy is required and how it
should be distributed in redshift and as a function of the local gas
density, to prevent overcooling and, at the same time, to reproduce
$X$--ray observables of galaxy clusters and groups.  A summary of the
characteristics of the heating recipes that we explore here is
provided in Table \ref{t:heat}. In the rest of the paper we will
present results based only on the high--resolution (HR) runs.

\subsubsection{Impulsive heating}
In our first class of heating schemes, we assume that all the energy
is dumped into the diffuse baryons in an impulsive way, with a single
heating episode occurring at some redshift $z_h$.
\begin{description}
\item[(a)] Entropy floors $S_{\rm fl}=50$ keV cm$^2$ at redshift
$z_h=9$ (S50-9 runs) and at $z_h=3$ (S50-3 runs), and $S_{\rm fl}=25$
keV cm$^2$ at $z_h=9$. In this scheme, the entropy associated with each
gas particle, $s=T/n_e^{2/3}$ ($T$: temperature in keV; $n_e$ electron
number density in cm$^{-3}$), is either increased to the value $S_{\rm
fl}$ if smaller than that, or otherwise left unchanged (see also
Navarro et al. 1995; Bialek et al. 2001; BGW).  The choice of $z_h=9$
corresponds to a heating epoch well before a substantial amount of gas in
simulations cools and forms stars and, therefore, heavily suppresses star
formation. The existence of a pristine SN generation (from the
so--called Pop III stars) has been invoked to account for the IGM
metal enrichment (e.g., Madau, Ferrara \& Rees 2001). However, were
this heating able to rise the entropy to the above levels, it would
prevent the later formation of the Ly-$\alpha$ forest, which is known
to have about one order of magnitude lower entropy. Furthermore, the
amount of heating energy would also correspond to a too high
production of heavy elements. For these reasons, we consider this
choice for $S_{\rm fl}$ to be motivated by the phenomenology of X--ray ICM
properties alone, rather than by expectations from star formation processes at
high redshift. 
\item[(b)] A fixed amount of heating energy per particle, $E_h=0.75$
keV/particle at $z_h=3$. This amount of energy is roughly the same
as the average specific energy dumped by the S50-3 scheme within the
halo virial radius (see Table \ref{t:heat}). Therefore, it allows to
check for differences induced in the final results by distributing the
same energy budget in a different way as a function of gas
density. The $z_h=3$ heating epoch is close to that at which the
star-formation rate within a proto--cluster region peaks (e.g., Menci
\& Cavaliere 2000; Bower et al. 2001; BGW). An energy
budget $E_h\sim 0.6$--0.8 keV/part has been also suggested by
Finoguenov et al. (2001a) to be consistent with the Si abundance
detected in groups and clusters.
\end{description}

\begin{figure*}
\centerline{
\hbox{
\psfig{file=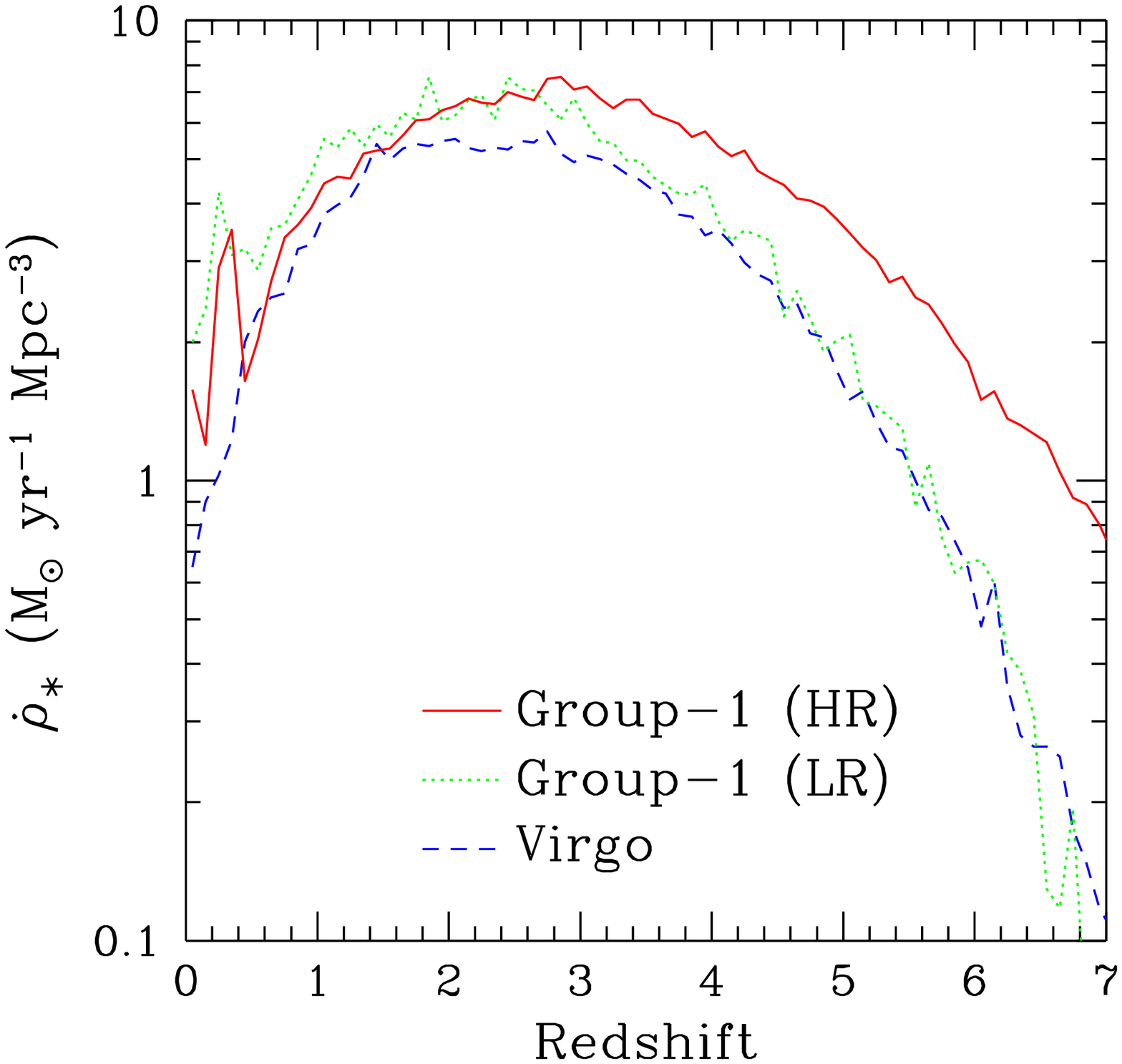,width=6cm} 
\psfig{file=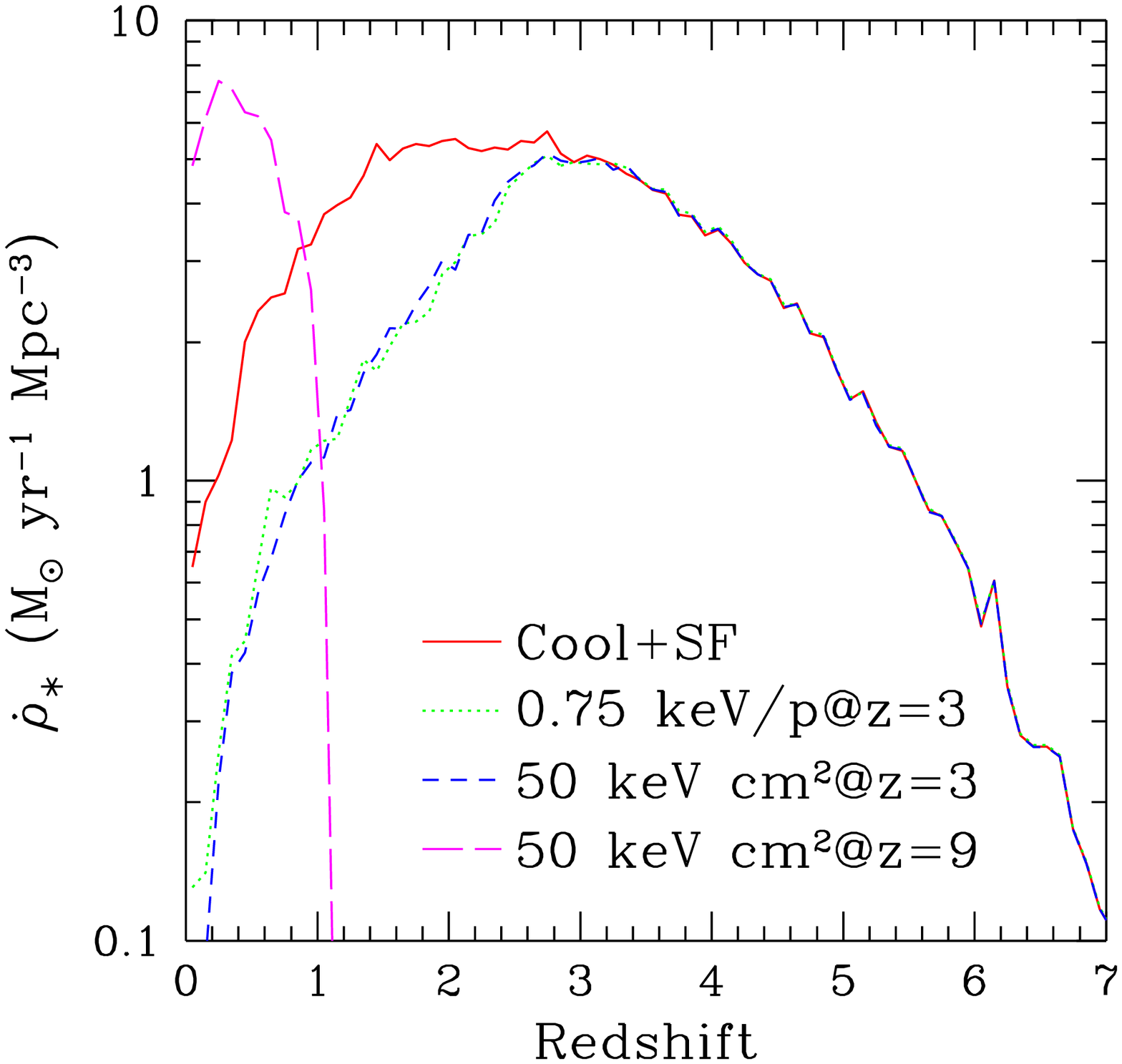,width=6cm}
\psfig{file=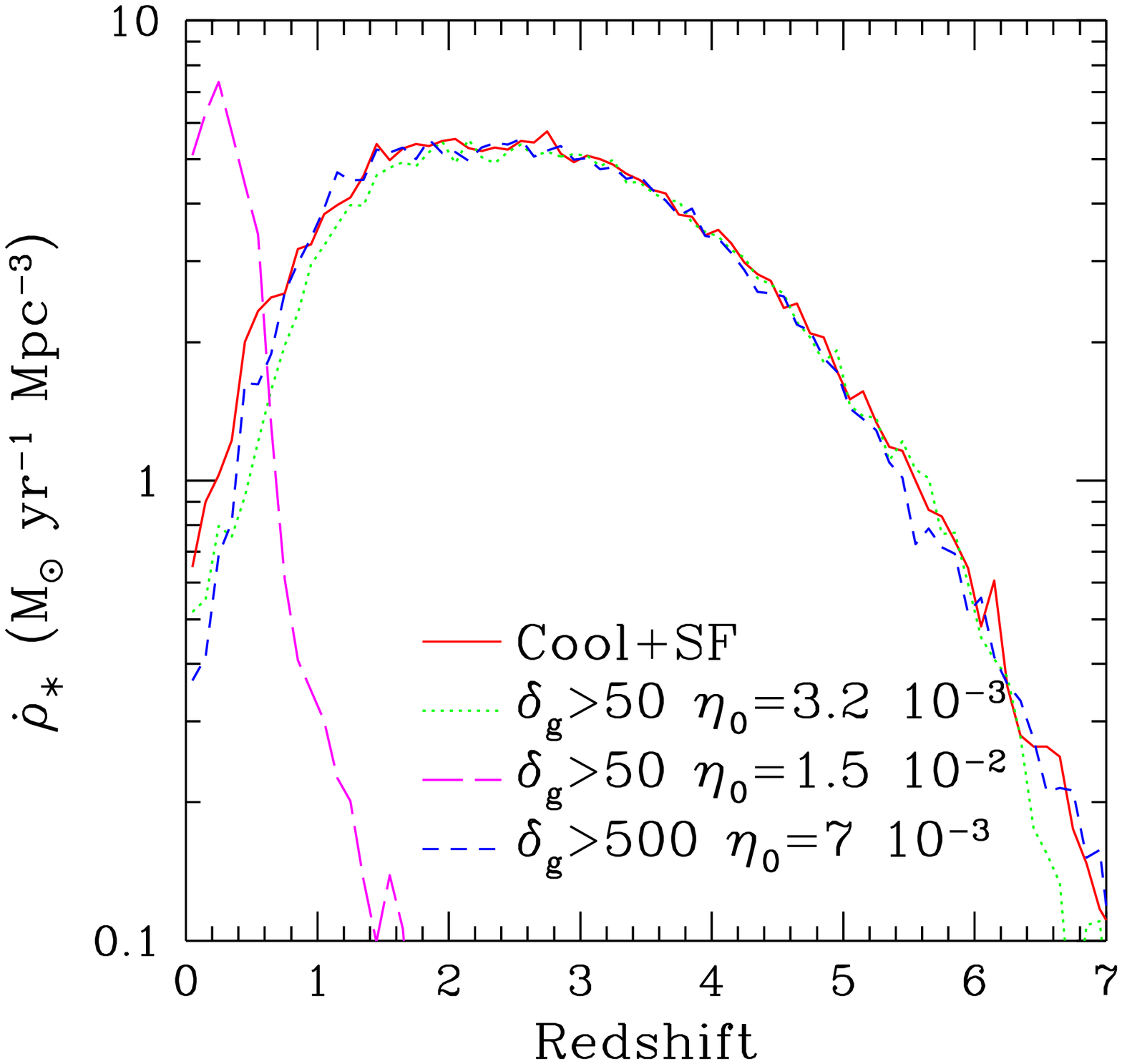,width=6cm}
}}
\caption{The density of the star--formation rate, $\dot\rho_\ast(t)$,
computed for the Lagrangian volume of the object that corresponds to the $z=0$
virial region of the simulated systems. Left panel: $\dot\rho_\ast(t)$
for the Virgo simulation (dashed curve), and for both the LR and HR
runs of Group-1 (dotted and solid curves, respectively), when no extra
heating is included. Central and right panels: results for the Virgo
cluster, simulated with impulsive heating and SAM--predicted SN
heating, respectively.}
\label{fi:sfr} 
\end{figure*}

\begin{table}
\centering
\caption{Prescriptions for non--gravitational heating. We give in
Column 1 name of the runs. For the impulsive heating schemes we give
in Column 2: the quantity which is modified by the heating; Column 3:
the heating redshift $z_h$. For the SAM--predicted SN feedback we give
in Column 2: the number of SN per unit $M_\odot$; Column 3: the
limiting density contrast for the gas particles to be heated. Column 4
gives the mean specific energy assigned to the gas particles falling
within $R_{\rm vir}$ by $z=0$. The asterisks indicate those runs which
have been realized only for the ``Virgo'' cluster.}
\begin{tabular}{lccc}
\multicolumn{4}{l}{Impulsive heating}\\
\multicolumn{1}{c}{Name of run} & \multicolumn{1}{c}{Scheme} & $z_h$ &
$E_h(<R_{\rm vir})$ \\
\hline
S25-9$^*$& $S_{\rm fl}=25$ keV cm$^2$ &   9   & 0.5   \\
S50-9    & $S_{\rm fl}=50$ keV cm$^2$ &   9   & 0.9   \\
S50-3    & $S_{\rm fl}=50$ keV cm$^2$ &   3   & 0.8   \\
K75-3    & $E_h=0.75$ keV/part&  3   & 0.75  \\
\hline \\
\multicolumn{4}{l}{SAM-predicted SN feedback}\\
\multicolumn{1}{c}{Name of run} & $\eta_0$ & $\delta_g$ & $E_h(<R_{\rm
  vir})$ \\ 
\hline
SN03$_L$      &  $3.2\,10^{-3}$ & 50  & 0.15 \\
SN07$_L$$^*$  &  $7.0\,10^{-3}$ & 50  & 0.32 \\
SN15$_L$$^*$  &  $1.5\,10^{-2}$ & 50  & 0.43 \\
SN07$_H$      &  $7.0\,10^{-3}$ & 500 & 0.36 \\
\hline
\end{tabular}
\label{t:heat}
\end{table}

\subsubsection{SAM--predicted SN feedback}
This heating scheme is based on computing the star--formation rate
(SFR) within clusters using a semi--analytic model (SAM) of
galaxy formation (e.g., Kauffmann, White \& Guiderdoni 1993,
Somerville \& Primack 1999, Cole et al. 2000, and references
therein). Here we employ a variation of the scheme described by Menci \&
Cavaliere (2000, see also Bower et al. 2001), and we refer to their
paper for a detailed description of the method, while we refer to BGW
for further details on the its implementation in cluster simulations.

The hierarchical merging of DM halos is followed by means of the
extended Press--Schechter formalism (e.g. Lacey \& Cole 1993), while
model parameters describing the gas physics, such as cooling, star
formation and stellar feedback, are chosen so as to reproduce observed
properties of the local galaxy population, such as the Tully--Fisher
relation, or optical luminosity functions and disk--sizes (e.g. Poli
et al. 2001). The model
prediction we are interested in here is the integrated star formation
history, $\dot m_*(z,M_0)$ ,of all the condensations which are
incorporated into a structure of total mass $M_0$ by the present
time. For a halo of size similar to our Virgo--like cluster, the SFR
peaks at $z\simeq 4$ in this semi-analytic model, while it is
$z\simeq2.5$--3 for the group--sized halos (see BGW, for a plot of the
$M_0$--dependence of the cluster SFR).

The rate of total energy feedback released by type-II SN is then
computed as
\be
{{\rm d}E_{SN}\over {\rm d}t}\,=\,10^{51}{\rm ergs}~\eta_0\,\dot m_*(z,M_0)\,,
\label{eq:snf}
\ee
where $\eta_0$ is the number of SN per solar mass of formed
stars. This value depends on the assumed initial mass function (IMF),
and is obtained by integrating over the IMF for stellar masses
$>8M_\odot$. In the following, we will use the values $\eta_0=3.2\times
10^{-3}M_\odot^{-1}$, which follows from a Scalo IMF (Scalo 1986),
$\eta_0=7\times 10^{-3}M_\odot^{-1}$, from the Salpeter IMF (Salpeter
1955) and $\eta_0=1.5\times 10^{-2}M_\odot^{-1}$, as an extreme case.

Since our simulations include cooling, radiative losses of SN energy
do not need to be assumed a priori, rather they are self--consistently
computed by the code. However, we need to specify the gas overdensity,
$\delta_g$, at which the SN heating energy is assigned to the gas. In
the following we take $\delta_{g}=50$ or 500, and assume that $E_{\rm
SN}$ is shared in equal parts among all the gas particles at
overdensity larger than $\delta_g$. The choice $\delta_g=50$
corresponds to assuming that the virial region of the whole halo is
heated and, therefore, that physical processes like galactic winds, for
example, are rather efficiently transferring energy to the
IGM. Increasing $\delta_g$ implies two competitive effects: on
one hand, it decreases the number of heated gas particles, therefore
it increases the amount of extra energy assigned to each of them; on
the other hand, the energy is assigned to denser particles, which have
shorter cooling time and, therefore, larger radiative losses.

\begin{figure*}
\centerline{
\hbox{
\psfig{file=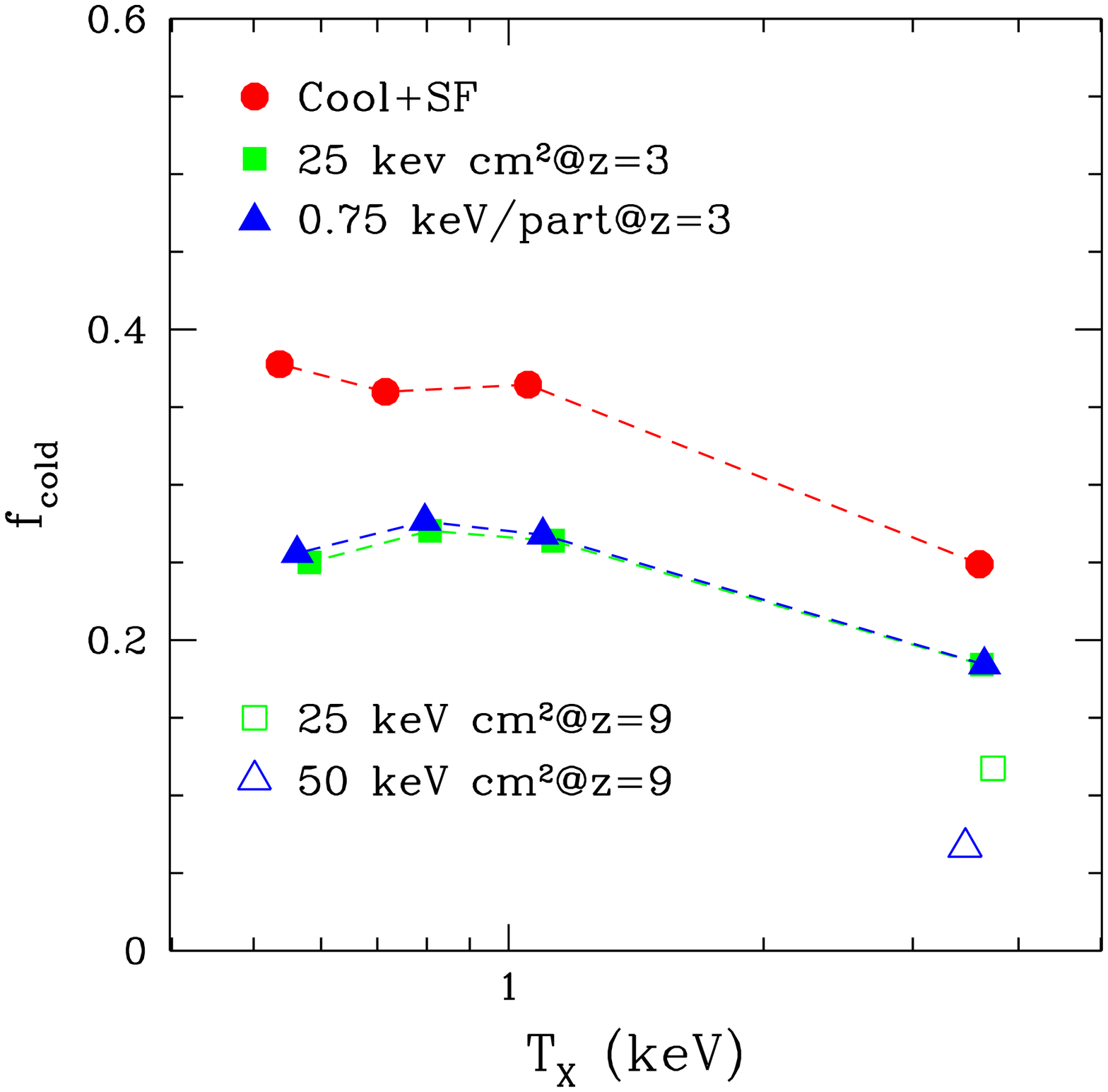,width=6.5cm} 
\psfig{file=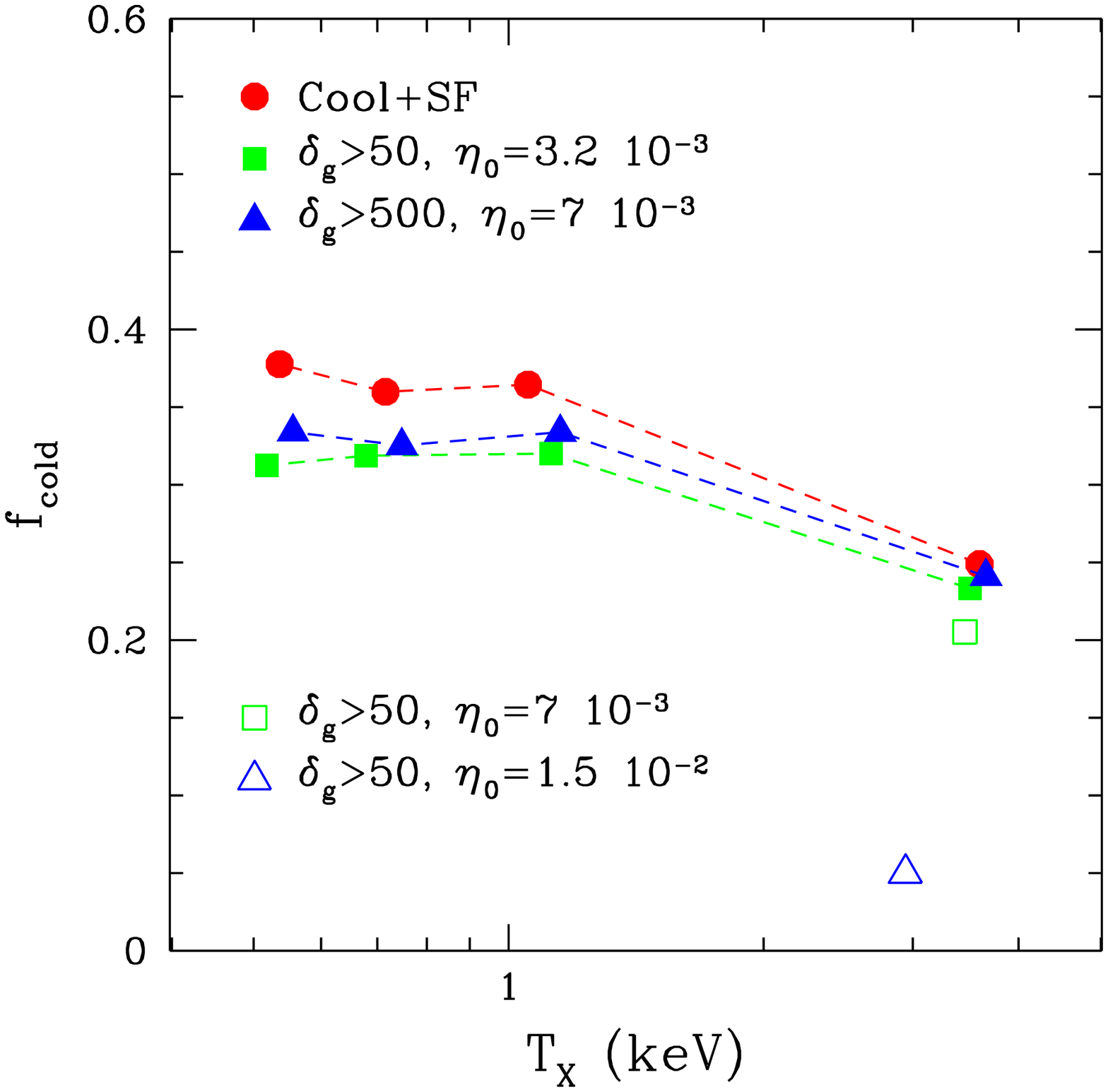,width=6.5cm}}}
\caption{The fraction of cold gas within the virial region of the
  simulated structures at $z=0$. Left and right panels show the effect
  of impulsive heating and of SAM-predicted SN feedback, respectively.}
\label{fi:fst} 
\end{figure*}

\subsection{The effect of extra heating on the cold fraction}
As we have already discussed, introducing cooling causes a too large
fraction of gas to be converted into stars. This is a well known
feature of hydrodynamical simulations, which has been widely discussed
in the literature (e.g., Suginohara \& Ostriker 1998, BGW). Even
worse, the runaway nature of the cooling process causes its efficiency
to be highly sensitive to numerical resolution (see
Fig.\ref{fi:fc_star}, see also Balogh et al. 2001). Therefore, one
should be very cautious in the interpretation of results from
simulations that do not resolve halos with luminosity well below $L_\ast$.

In Figure \ref{fi:fst}, we show the effect of the different heating
schemes on the resulting cold fraction within the virial radius of our
simulated structures. As expected, we find a decrease of $f_{\rm
cold}$ when non--gravitational heating is included. However, the
efficiency of this suppression of star--formation does not exclusively
depend on the amount of dumped energy. For instance, imposing an
entropy floor of 50 keV cm$^2$ at $z_h=9$ (left panel of
Fig.\ref{fi:fst}) is far more efficient than at $z_h=3$. The reason
for this is illustrated by the different patterns of SFR history, that
we show in Figure \ref{fi:sfr}. The impulsive heating at $z_h=3$
causes a suppression of the SFR at later epochs, but a fair amount of
stars are already in place at $z_h$ (central panel of
Fig. \ref{fi:sfr}). Quite interestingly, the results for the two runs
with heating at $z_h=3$ produce quite similar SFR. This indicates
that, once the heating epoch is fixed, the degree of SFR suppression
depends only on the amount of heating energy, while being largely
independent on its distribution as a function of the local gas
density.

However, heating at $z=9$ with a comparable amount of energy
does not allow gas to reach high densities within DM halos and to cool
before $z\simeq 1$. Once cooling takes place, it converts less than
$10$ per cent of the gas into stars, within a short episode of star
formation. The resulting SFR peaks at very low redshift, $z\simeq
0.3$, which is highly discrepant with observational determinations of
the SFR history in clusters (e.g., Kodama \& Bower 2001). Of course,
this is not the only feature which rules out the picture of a strong
heating occurring at such a high redshift. For instance, at $z=0$,
stars are all concentrated in one single object located at the center,
a cD--like galaxy, while no other galaxy--sized DM halos contains
significant amounts of collapsed gas.

\begin{figure*}
\centerline{
\psfig{file=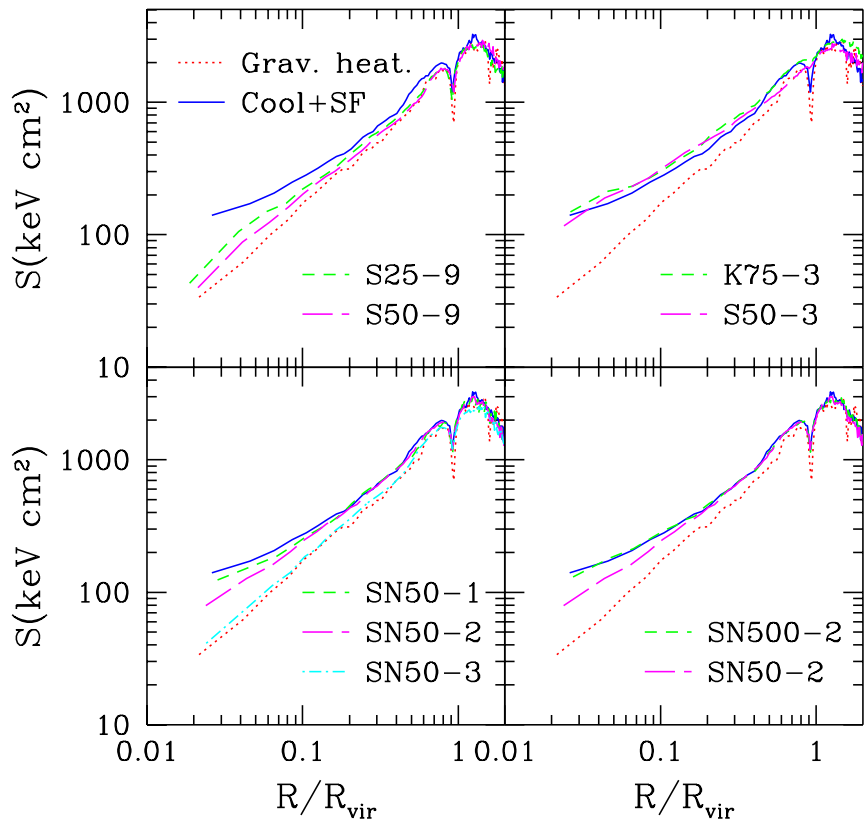,width=12cm} 
}
\vspace{-0.5truecm}
\caption{The entropy profiles of the ``Virgo'' cluster
simulations. The upper panels show the effect of impulsive heating at
$z=9$ (on the left) and $z=3$ (on the right), while the lower panels
show the effect of SN feedback from the SAM--predicted SFR history,
after changing $\eta_0$ (on the left) and the density threshold for
gas heating (on the right). For reference, the entropy profiles for
the run with gravitational heating and for the run including only
cooling and star formation are shown in all the panels with the dotted
and the solid lines, respectively.}
\label{fi:sprof}
\end{figure*}

As for the SN heating (right panel of Fig. \ref{fi:sfr}), suppressing
the star fraction below the 10 per cent level requires a high,
probably unrealistic value for $\eta_0$. Also, taking
$\eta_0=1.5\times 10^{-2}M_\odot^{-1}$ generates an implausible SFR
history, resembling that found for the runs based on setting the
entropy floor at $z_h=9$: the large amount of extra heating at high
redshift prevents the occurrence of star formation down to $z\simeq
1.5$. Taking $\eta_0$ in the range 3--7$\times 10^{-3}$ predicts more
realistic SFRs, but it is not able to suppress $f_{\rm cold}$ below
the $\simeq 20$ per cent level.

A general conclusion of our analysis is that heating schemes producing 
plausible SFR histories are not efficient in suppressing the fraction 
of cold gas below the 20 and 25 per cent values at the cluster and group 
scales, respectively. Vice-versa, a more efficient suppression is obtained 
by preventing gas to cool at high redshift, at the expense of delaying
star formation to unreasonably low redshifts.

\section{$X$--ray properties of simulated clusters}
\subsection{The entropy of the ICM}
Measurements of the excess entropy in central regions of poor clusters
and groups are considered to provide direct evidence for the lack of
self--similarity of the ICM properties (e.g., Ponman et al. 1999;
Finoguenov et al. 2002a). In a separate paper (Finoguenov et
al. 2002b), a self--consistent comparison is realized between the
entropy properties of the simulations with impulsive heating, that we
present here, and the observational data for groups and clusters by
Finoguenov et al. (2002a). The main result of this comparison is that,
although cooling and star formation tend to somewhat increase entropy
in central cluster regions, they still fall short in producing the
entropy excess which is observed at the group scale. While preheating
at $z_h=3$ is shown to increase the entropy to the observed values,
runs with $z_h=9$ are characterized by a low entropy level in central
regions of clusters and groups.

\begin{figure*}
\centerline{\vbox{
\hbox{
\psfig{file=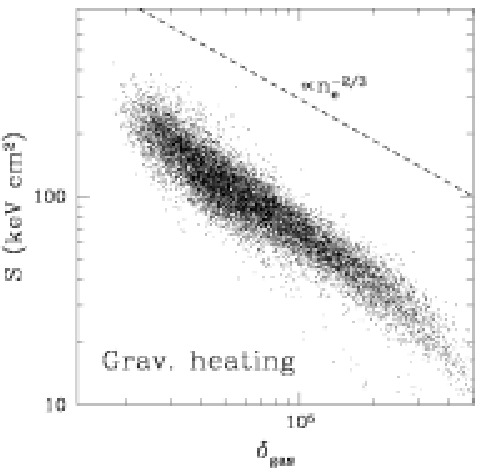,width=6.cm} 
\psfig{file=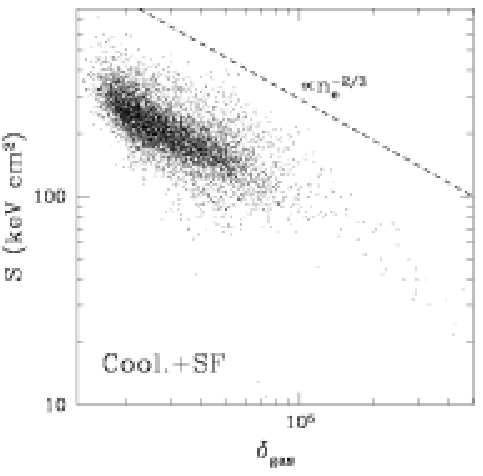,width=6.cm}
}
\vspace{0.1cm}
\hbox{
\psfig{file=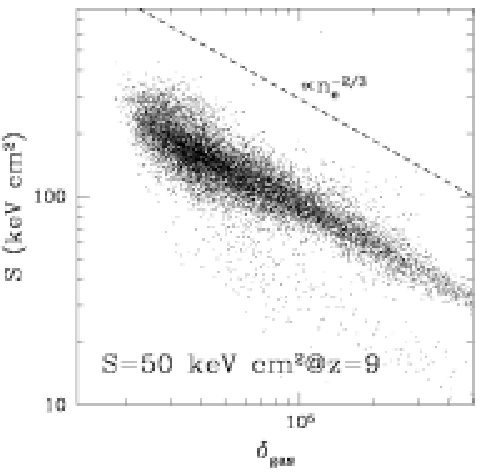,width=6.cm} 
\psfig{file=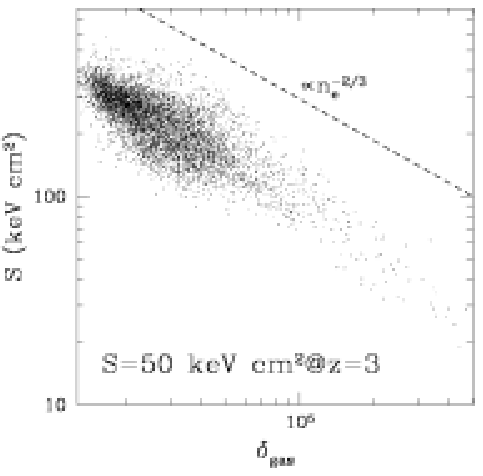,width=6.cm}
}
}}
\caption{The entropy--overdensity phase--diagram for the gas
particles falling within $0.1\,R_{\rm vir}$ at $z=0$, for different runs of
the ``Virgo'' cluster. In each panel, the dashed line shows the scaling
$S\propto n_e^{2/3}$ expected for isothermal gas.}
\label{fi:srho} 
\end{figure*}

Instead of attempting any further comparison with observations, we
want to discuss here the dynamical reasons for such a behavior.  To
this end, we show in Figure \ref{fi:sprof} the effect of cooling
and non--gravitational heating on the entropy profiles for our whole
set of ``Virgo'' simulations.  As expected, when cooling and SF are
included, low entropy gas is selectively removed in central cluster
regions, thus inducing a flattening of the profile. This is explicitly
shown in Figure \ref{fi:srho}: while the run including only
gravitational heating has a population of high--density low entropy
gas particles, such particles are removed from the diffuse phase once
cooling and star formation are introduced. This result is consistent
with the expectation from analytical arguments based on the comparison
between cooling time--scale and typical cluster age (e.g., Voit et
al. 2002, Wu \& Xue 2002). The inclusion of extra heating has a
non--trivial effect on the efficiency of cooling in removing particles
from the lower left side of the $S$--$\delta_g$ phase diagram.  For
instance, imposing the same entropy floor at $z_h=9$ and at $z_h=3$
has quite different effects on the entropy pattern (see lower panels
of Fig. \ref{fi:srho}). Heating at $z_h=9$ has the effect of
increasing the cooling time for most of the gas particles, so as to
allow star formation to take place only quite recently (see
Fig. \ref{fi:sfr}). The increased time-scale for cooling causes this
process to proceed in a more gradual way. For this reason, the entropy
of gas particles undergoing cooling decreases slowly, thus making
their removal from the hot diffuse phase less efficient.

\subsection{The luminosity--temperature and luminosity--mass relations}
The slope of the $L_X$--$T$ relation also provides important
observational evidence for the lack of self--similar behaviour of the
ICM. Since the first measurements of ICM temperatures for sizable sets
of clusters, it has been recognised that $L_X\propto T^\alpha$ with
$\alpha\simeq 3$, although with a considerable scatter (e.g., White et
al. 1998, and references therein). Better quality observations
established that a significant contribution to this scatter is
associated with the different strength of cooling flows detected in
different clusters. Either excluding clusters with pronounced
signatures of cooling flows or correcting for their effect (e.g.,
Markevitch 1998; Allen \& Fabian 1998; Arnaud \& Evrard 1999; Ettori
et al. 2002) results in a much tighter $L_X$--$T$ relation, albeit still with
a rather steep slope. At the same time, hints have also been found for
a further steepening of this relation at $T\mincir 1$ keV (e.g.,
Helsdon \& Ponman 2000, and references therein), possibly indicating
that the mechanism responsible for the $L_X$--$T$ scaling should act
in a different way for clusters and groups.

Simulations that allow for non--gravitational heating (e.g., Bialek et
al. 2001; BGW) and radiative cooling (e.g. Pearce et al. 2001; Dav\`e
et al. 2002; Muanwong et al. 2002) have been shown to be able to
account for the observed $L_X$--$T$ relation. However, a sometimes
overlooked issue in determining the $X$--ray luminosity of clusters in
simulations concerns the contribution of metal lines to the
emissivity. While this contribution is negligible above 2~keV, it
becomes relevant at the scale of groups. For instance, neglecting the
contribution from line emissivity for an ICM enriched to a metallicity
of $Z=0.3Z_\odot$ leads to an underestimate of the X--ray luminosity by
almost 50 per cent at 1 keV, and by more than a factor 2 at 0.5 keV
(e.g. BGW).

A correct procedure would require simulations that include a treatment
of metal enrichment and a self--consistent estimate of the
contribution of line cooling to the X--ray emissivity. However, only
preliminary attempts have been realized so far to include the
treatment of ICM metal enrichment from SN ejecta (Lia et al. 2002;
Valdarnini 2002). Pearce et al. (2001) include the contribution of
metals to the cooling function adopted in their simulation by assuming
$Z=0.3\,Z_\odot$ at the present epoch, linearly decreasing with time
towards the past. Dav\'e et al. (2002) did not include the metal
contribution in their cooling function, but estimated $X$--ray
luminosities by assuming a phenomenological relation between metal
abundance and temperature of the galaxy system. However, while the ICM
metallicity at the scale of rich clusters is quite well established
from observations, the situation is less clear for poor clusters and
groups (e.g., Davis, Mulchaey \& Mushotzky 1999; Renzini 2000, and
references therein).

The cooling function used in our simulations assumes zero metallicity,
but we compute the $X$--ray luminosity by adding to the bremsstrahlung
emissivity the contribution from lines for a $Z=0.3\,Z_\odot$
plasma. This represents a reasonable approximation as long as gas
spends most of the time at low metallicity, being enriched to high
metallicity only recently. Owing to the uncertainties connected to
these assumptions, the reliability of $L_X$ values at $T\mincir 1$ keV
is unclear, however. Precise predictions will require a fully
self--consistent treatment of metal enrichment of the ICM from star
formation activity.

\begin{figure*}
\centerline{
\psfig{file=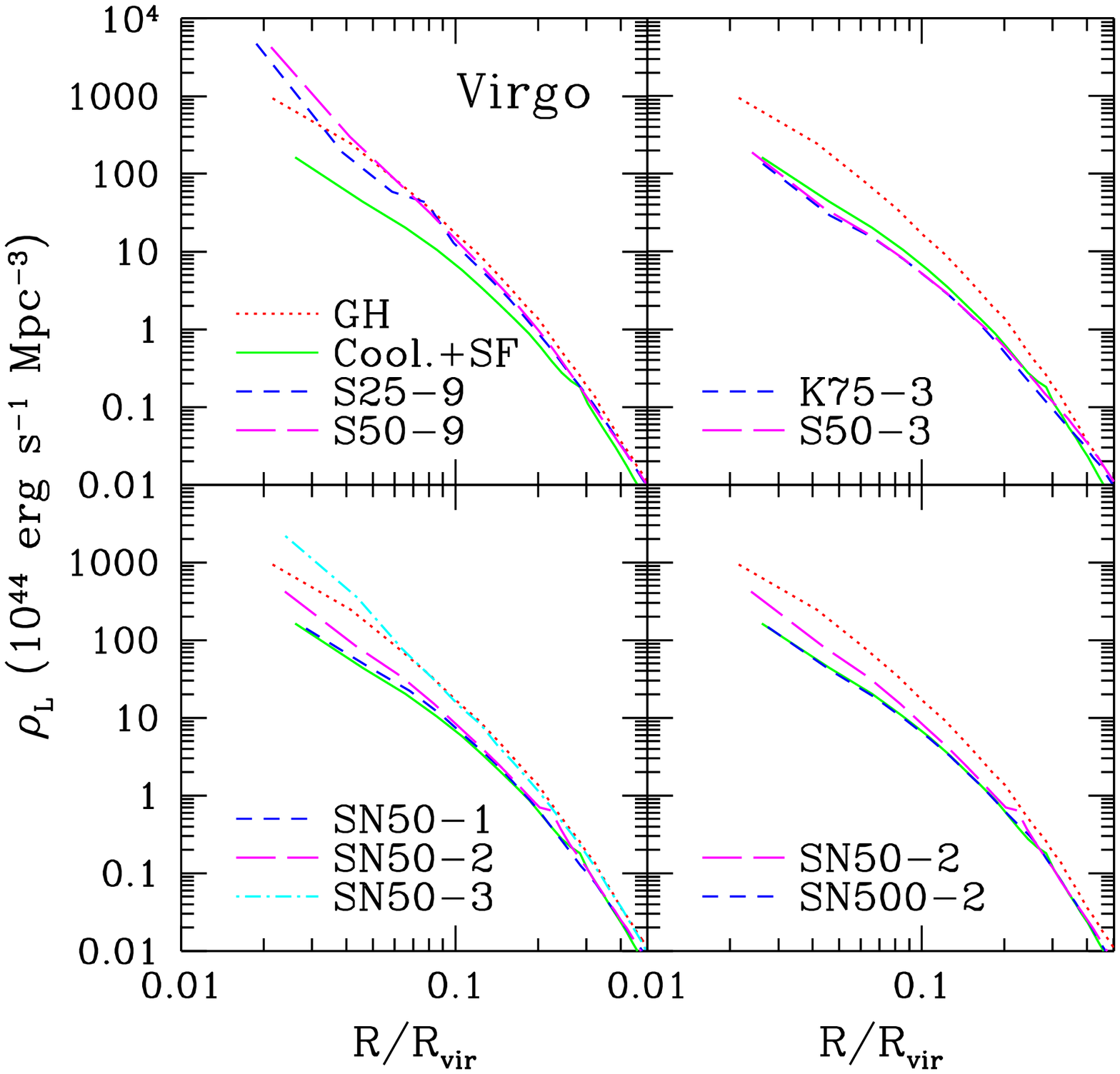,width=12cm} 
}
\vspace{-0.5truecm}
\caption{The profiles of $X$--ray luminosity density for the ``Virgo''
runs. The sequence of panels is the same as in Figure \ref{fi:sprof}.}
\label{fi:lprof}
\end{figure*}

In Figure \ref{fi:lprof}, we show the profiles of emissivity (energy
released per unit time and unit volume) for the different Virgo
runs. As expected, including only cooling and star formation has the
effect of flattening the profiles in the central cluster regions as a
consequence of gas removal from the hot phase. When extra heating is
included, the profiles change according to the amount of gas left at
relatively low entropy in the central cluster regions. For instance, the
fairly large population of low entropy particles in the run with
$S_{\rm fl}=50$ keV cm$^2$ at $z_h=9$ (see Fig.\ref{fi:srho}) is
responsible for the spike in the $X$--ray emissivity. In the same way,
the efficient removal of low--entropy gas for the run where an entropy
floor was imposed at $z_h=3$ explains the flattening of the luminosity
profile in the central cluster region.  These results confirm the
existence of a non--trivial interplay between the effects of cooling
and extra heating. In some cases, one reaches the apparently
paradoxical conclusion that combining heating and cooling increases
the $X$--ray luminosity, although their separate effects are that of
suppressing $L_X$.

\begin{figure*}
\centerline{
\hbox{
\psfig{file=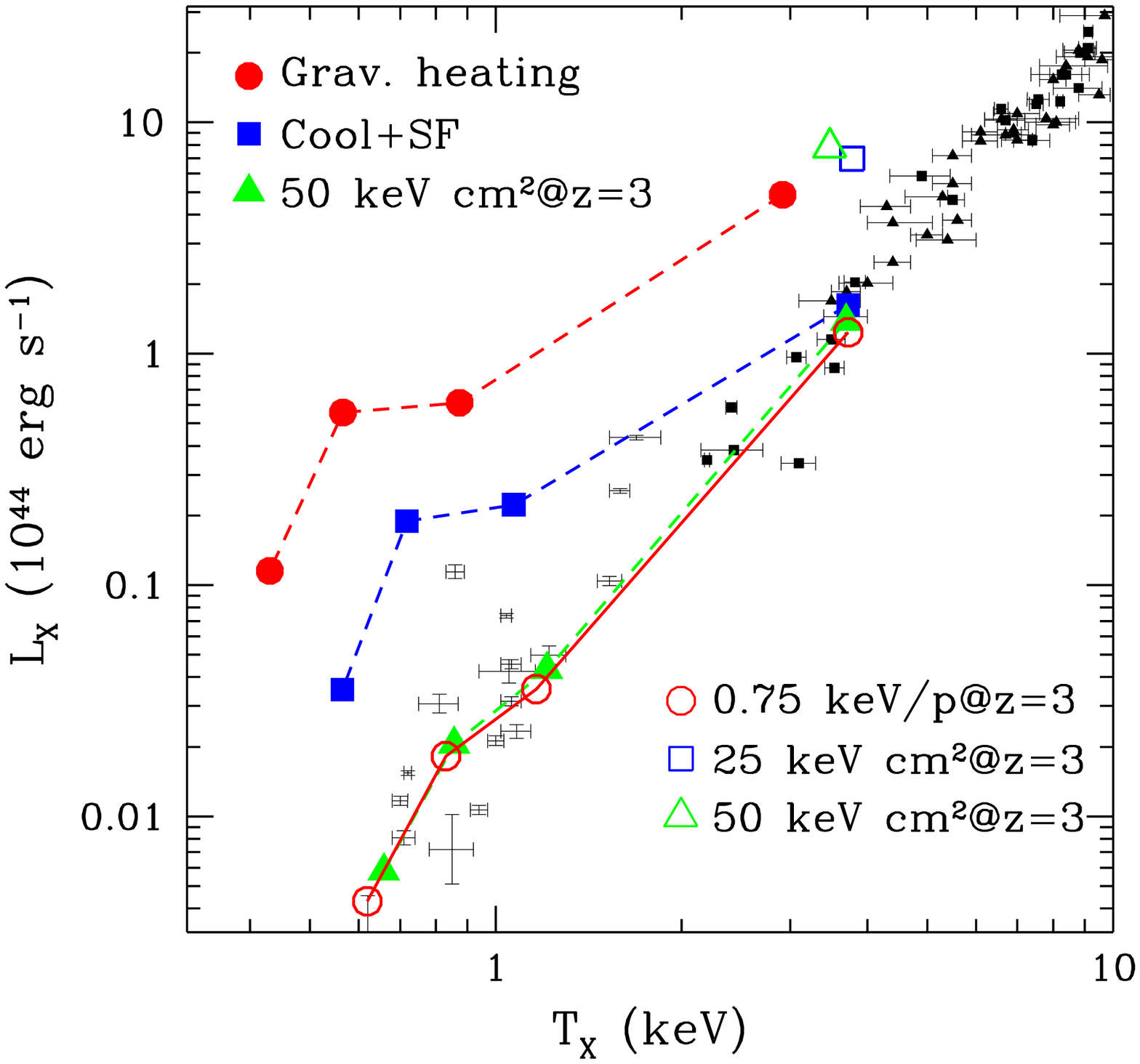,width=6.5cm} 
\psfig{file=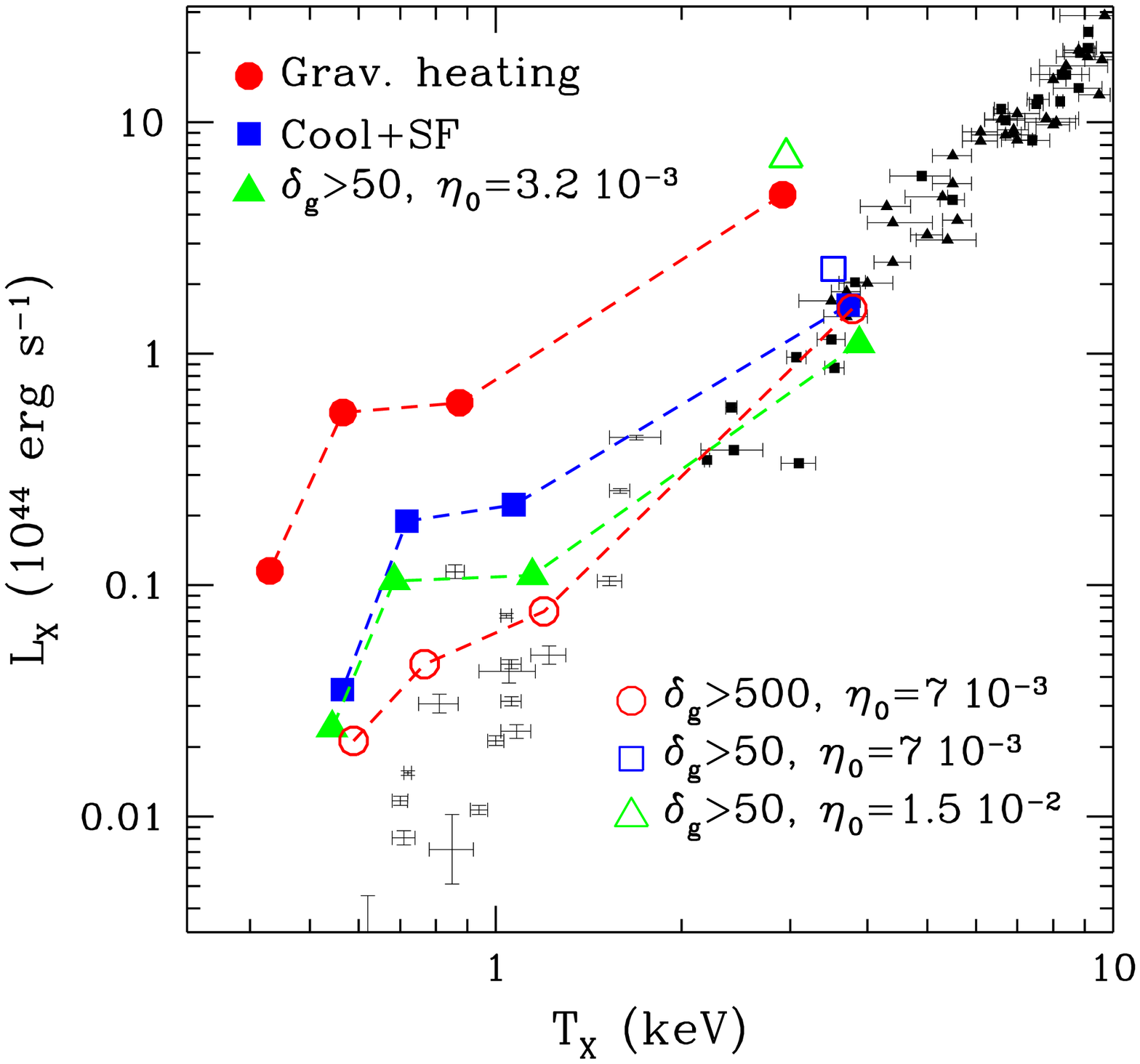,width=6.5cm}}}
\caption{The relation between bolometric luminosity and emission
weighted--temperature for the simulations and for observational data
at $z=0$. The right and the left panel are for the effects of impulsive
heating and SN feedback, respectively. Data points at the cluster
scale are from Markevitch (1988, small triangles) and from Arnaud \&
Evrard (2000, small squares), while data for groups (crosses) are from
Helsdon \& Ponman (2000).}
\label{fi:l-t} 
\end{figure*}

Figure \ref{fi:l-t} shows the comparison between the simulated and the
observed $L_X$--$T$ relation for clusters and groups. As expected,
cooling causes a sizeable suppression of the $X$--ray luminosity. At
the same time, the emission--weighted temperature is increased as a
consequence of the steepening of the temperature profiles in the
central halo regions (see below). While the mere introduction of
cooling and star--formation brings the ``Virgo'' cluster into
agreement with observations, the simulated groups are somewhat
overluminous with respect to data. The inclusion of pre--heating at
$z_h=3$ has a smaller effect on the Virgo cluster, consistent with the
result from the luminosity profile, while it further suppresses $L_X$ at
the scale of groups. A similar result is also found for the runs with
SAM--predicted SN feedback.

Quite interestingly, the $L_X$ value for Group-2 in the runs with
no extra heating appears to be systematically in excess with respect to
that inferred from the $L_X$--$T$ scaling of the other three simulated
structures. This deviation is due to the occurrence of a recent merger
shock in the Group-2 run, which produced a sudden increase in the
$X$--ray emission. When extra heating is included, its effect is that
of decreasing the strength of the shock, thus also reducing the jump
in luminosity.

\begin{figure}
\centerline{
\psfig{file=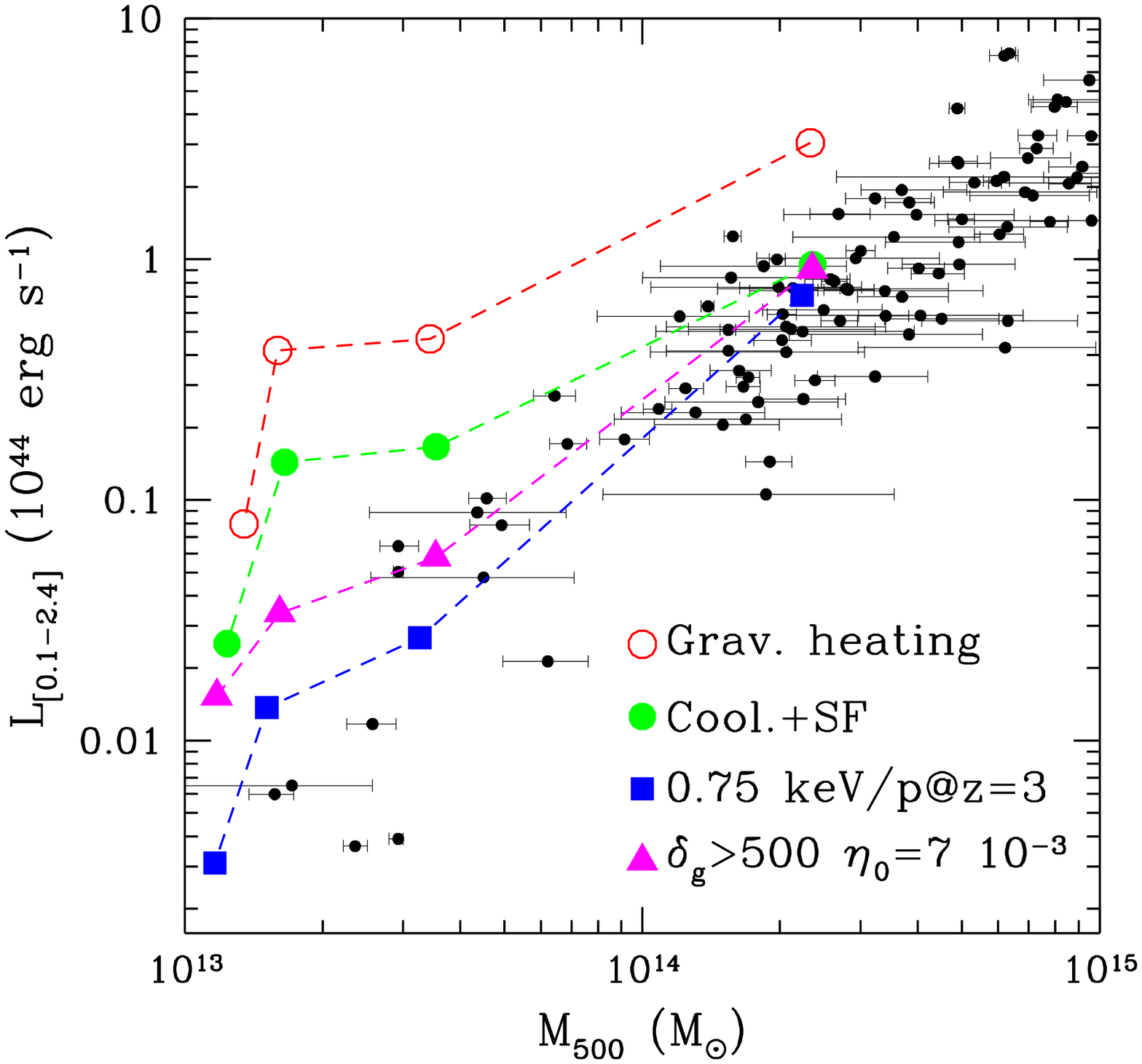,width=8.5cm} }
\caption{Comparison of simulations and observational results for the
relation between X--ray luminosity in the 0.1--2.4 keV energy band and
the mass at an overdensity $\bar \rho/\rho_{\rm crit}=500$. Small
circles with errorbars are the observational data points from Reiprich
\& B\"ohringer (2002).}
\label{fi:lm} 
\end{figure}

A similar constraint is provided by the relation between $X$--ray
luminosity and mass. Reiprich \& B\"ohringer (2002) have estimated
this relation by applying the equation of hydrostatic equilibrium to a
fairly large ensemble of clusters and groups, under the assumption of
isothermal gas. They used ICM temperatures based on ASCA data, in
combination with ROSAT-PSPC data for the surface brightness
profile. Ettori, De Grandi \& Molendi (2002) used the better quality
data from Beppo--SAX observations to resolve the temperature profiles
for a smaller ensemble of clusters. Although the analysis by Ettori et
al. explicitly includes temperature gradients when solving the equation
of hydrostatic equilibrium, it is restricted to clusters with
$T\magcir 3$ keV, thus hotter than those simulated here. For this
reason, we here compare our simulation results to the data by
Reiprich \& B\"ohringer (see Figure \ref{fi:lm}). In this analysis,
the cluster masses, $M_{500}$, are computed within the radius encompassing
an average density $\bar\rho=500\rho_{\rm crit}$, while observed
luminosities are provided in the 0.1--2.4 keV ROSAT energy band. We
use the MEKAL spectral model to correct bolometric luminosities from
simulations by assuming $Z=0.3\,Z_\odot$ for the global ICM
metallicity. Consistent with the results from the analysis of the
$L_X$--$T$ relation, we find that the runs with heating at $z_h=3$ and
that with SN feedback, based on a Salpeter IMF, are able to follow the
steep slope of the observed $L_{0.1-2.4}$--$M_{500}$ relation.

In principle, the $L_X$--$T$ and the $L_X$--$M$ relations do not
provide independent information, since masses are anyway estimated
using temperature data. Still, both relations are obtained by using
largely different observational data sets and analysis
procedures. Therefore, the fact that the same simulations are able to
account for both scalings lends support to the robustness of our
results and indicates that our conclusions are not affected by
observational biases or systematics.

Owing to the uncertainties mentioned above in modelling the
luminosities of groups, it appears prudent not to make strong claims
about how much extra heating is needed to reproduce the
observations. Overall, we note that all the runs that produce a
delayed star formation, such as those with $z_h=9$ and the one with
SN--feedback and large $\eta_0=1.5\times 10^{-2}M_\odot^{-1}$ (see
Fig.\ref{fi:sfr}), are quite inefficient in suppressing $L_X$. On the
contrary, runs with pre--heating at $z_h=3$ or with SN--feedback
combined with more reasonable values for $\eta_0$ succeed to account
for the steep slopes of the $L_X$--$T$ and $L_X$--$M$ relations.

Having warned about the reliability of the emissivity modeling for gas
at $T<1$ keV, a word of caution should also be spent on the
reliability of the interpretation of current observational
data. Estimating temperature and luminosity for small groups from
pre--Chandra and pre--XMM data is not a trivial task, mostly due to
the difficulty of separating the contribution of the diffuse intra--group
medium from that of member galaxies, and of detecting $X$--ray
emission out to a significant fraction of the virial radius (see,
e.g., Mulchaey 2000, for a review on the $X$--ray properties of
groups). The situation is likely to improve as newer and better
quality data will be accumulated, although we will probably have to
wait for a few more years before a critical amount of Chandra and
Newton--XMM observations of groups will be available.

\begin{figure*}
\centerline{
\hbox{
\psfig{file=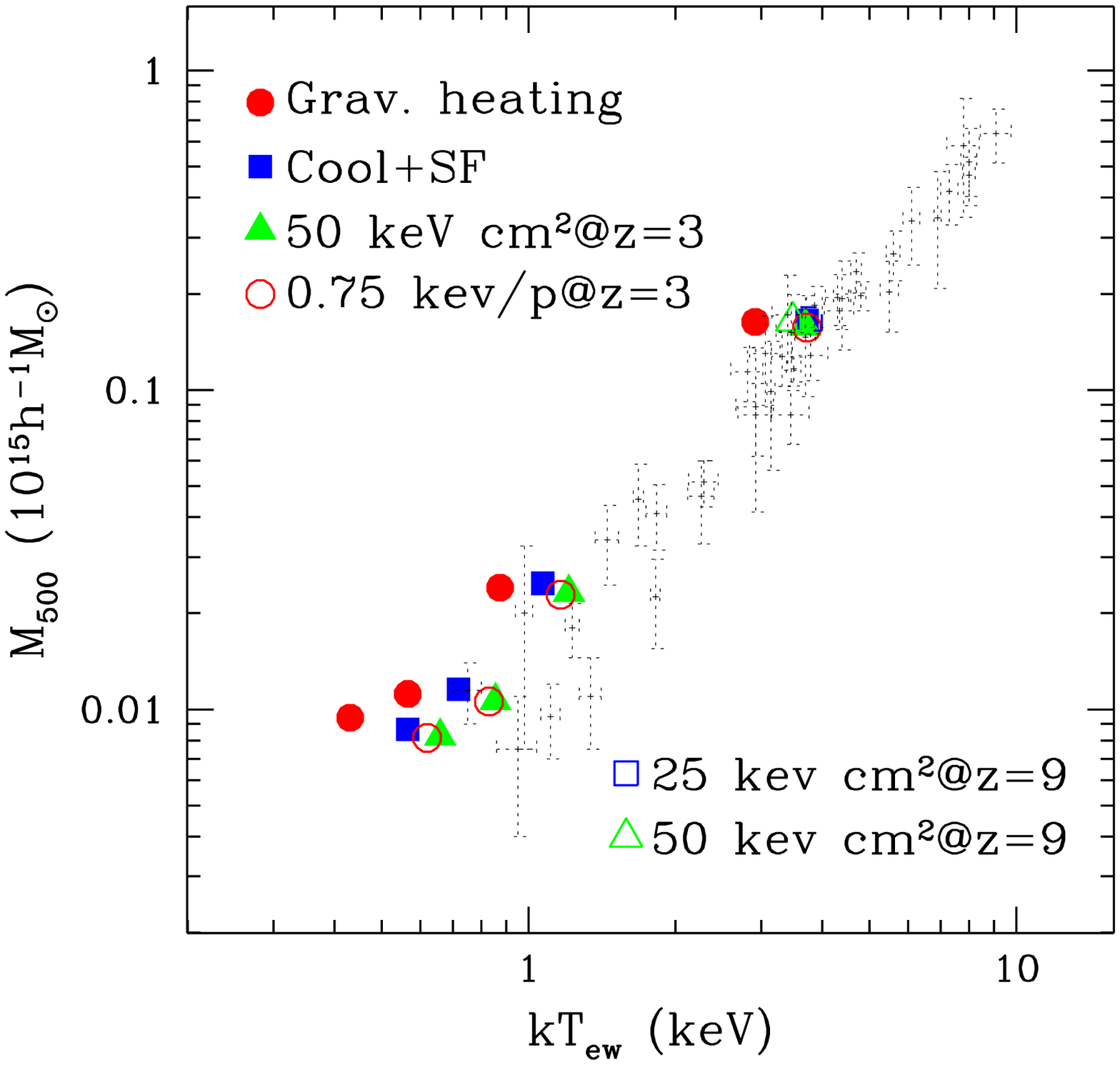,width=6.5cm} 
\psfig{file=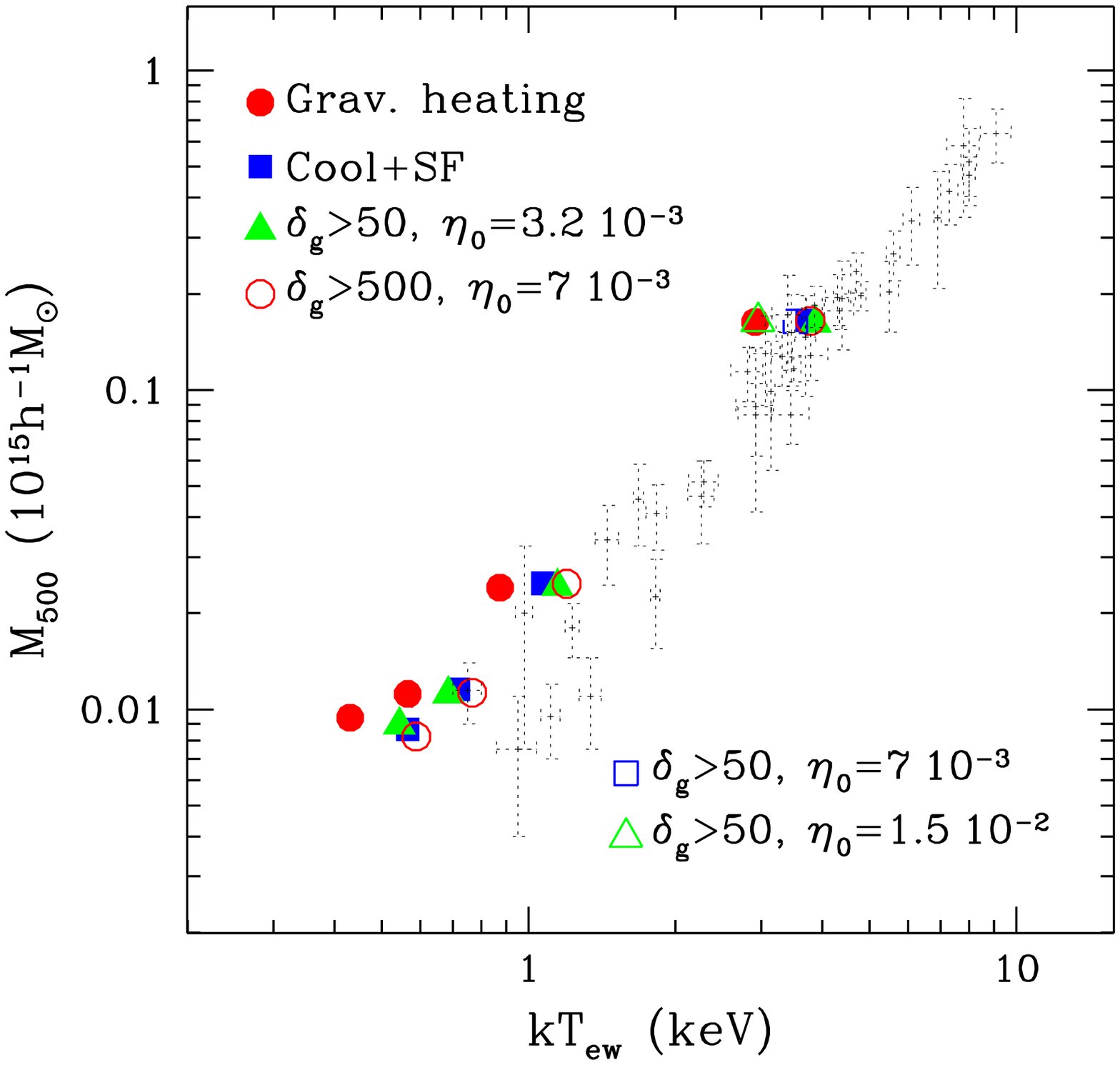,width=6.5cm}}}
\caption{The relation between the mass at overdensity $\rho
/\rho_{\rm crit}=500$ and the emission weighted--temperature. The
right and the left panels are for the effects of impulsive heating and
SN feedback, respectively. Data points are from Finoguenov et
al. (2001b).}
\label{fi:m-t}
\end{figure*}

\subsection{The mass--temperature relation}
Under the assumptions of spherical symmetry and an isothermal gas
distribution, the condition of hydrostatic equilibrium predicts a
precise relationship between the virial mass of a cluster and its
temperature:
$k_BT=1.38\beta^{-1}M_{15}^{2/3}[\Omega_m\Delta_{\rm vir}(z)]^{1/3}(1+z)$
keV for a gas of primordial composition, with $M_{15}$ being the virial mass
in units of $10^{15}h^{-1}M_\odot$ and $\Delta_{\rm vir}$ being the ratio
between the virial density and the average cosmic matter density at
redshift $z$. Under the above assumptions, the $\beta$ parameter gives
the ratio between the specific kinetic energy of dark matter particles
and the thermal energy of the gas. Simulations including only
gravitational heating demonstrated that this relation is reproduced
quite well, with $\beta\simeq 1$--1.2 (e.g., Evrard et al. 1996; Bryan
\& Norman 1998; Frenk et al. 1999; BGW). For these reasons,
the $M$--$T$ relation has been considered for several years as a
fairly robust prediction of hydrostatic equilibrium: gas temperature,
unlike $X$--ray emissivity, is primarily determined by the action of
gravity and, as such, depends on global cluster properties, and only
weakly on local structure of the ICM.

However, data based on ASCA and ROSAT observations show an $M$--$T$
relation which is about 40 per cent lower than predicted (Horner et
al. 1999; Nevalainen et al. 1999; Finoguenov et al. 2001b), a result
which has been confirmed by Beppo--SAX (Ettori, De Grandi \& Molendi
2002) and Chandra (Allen et al. 2001) data for relatively
hot systems ($T\magcir 4$ keV).

Non--gravitational heating could be naively expected to solve this
discrepancy by increasing the ICM temperature at fixed cluster
mass. However, BGW have shown that for a broad class of pre--heating
models similar to those discussed here the $M$--$T$ relation is left
almost unchanged by the injection of extra--energy (cf. also Lin et
al. 2002). In fact, as long as gas has time after being heated to
settle back into hydrostatic equilibrium within the gravitational
potential well, its temperature is mainly determined by the amount of
collapsed dark matter, which is unaffected by the heating process.

An alternative explanation for the observed low amplitude of the
$M$--$T$ relation, based on the effect of radiative cooling, has been
shown by Thomas et al. (2002) to be much more promising. In this case,
gas left in the diffuse phase flows towards the central cluster
region, where it is compressed, thus increasing its temperature. As a
result, the overall mass--weighted temperature remains almost
unchanged, but the emission weighted temperature significantly
increases. Our results, as shown in Figure \ref{fi:m-t}, actually
confirm this picture and generalise it to a large range of schemes for
extra heating: while the value of $M_{500}$ is left
unchanged by the cooling/heating processes, $T_{\rm ew}$ increases as a
consequence of the temperature increase in the central cluster regions.

\begin{figure}
\centerline{
\psfig{file=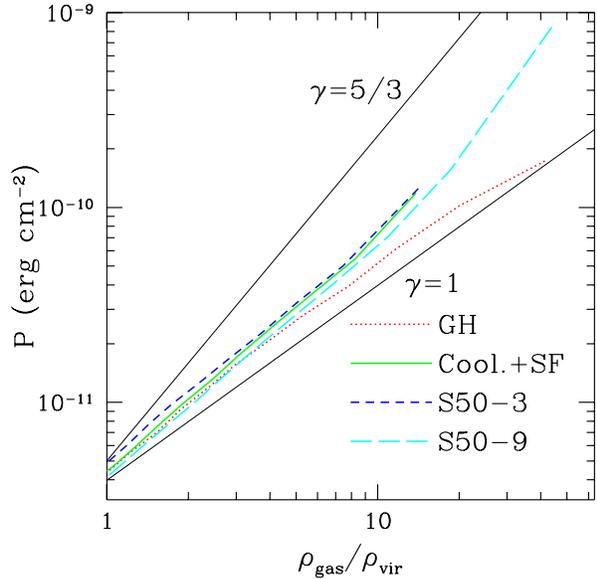,width=8.5cm} 
}
\vspace{-0.5truecm}
\caption{The relation between gas pressure (cgs units) and density (in
  units of the average total density within the virial radius),
  computed within spherical shells. The two thin solid lines
  correspond to effective polytropic indices $\gamma= {\rm d}\log P/
  {\rm d}\log \rho_{gas} =1$ (isothermal model) and $\gamma=5/3$
  (adiabatic model).}
\label{fi:gamma}
\end{figure}

In order to better understand the effect of cooling on the central
temperature structure of the ICM, we plot in Figure \ref{fi:gamma} the
gas pressure, $P=\rho_{\rm gas}k_BT/(\mu m_p)$, as a function of gas
density for the simulations of the ``Virgo'' cluster. We introduce
here the effective polytropic index $\gamma ={\rm d}\log P/{\rm d}\log
\rho_{\rm gas}$ to describe the run of pressure as a function of gas
density. In the external cluster region the gas is characterised by
$\gamma\magcir 1$, thus consistent with the slowly outward-declining
temperature profiles, almost independent of the presence of cooling
and extra heating. However, cooling leads to a loss of pressure in
central high--density regions. As cooling partially removes
low--entropy gas from the diffuse phase, gas of higher entropy flows
in from more external regions. As long as this gas has sufficiently
long cooling time, its entropy is conserved and the gas is
adiabatically compressed during the inflow.  In this regime, the
effective polytropic index increases towards $\gamma=5/3$, thus
indicating an adiabatic behaviour of the ICM. This result is
essentially independent of whether gas is preheated. The only effect
of imposing an entropy floor at $z_h=9$ is that of making the cooling
process more gradual. This allows a larger amount of gas to remain in
the diffuse phase, so as to reach higher density and higher pressure
in central regions (see also Fig.~\ref{fi:lprof}).

\subsection{The temperature profiles}
\begin{figure*}
\centerline{
\psfig{file=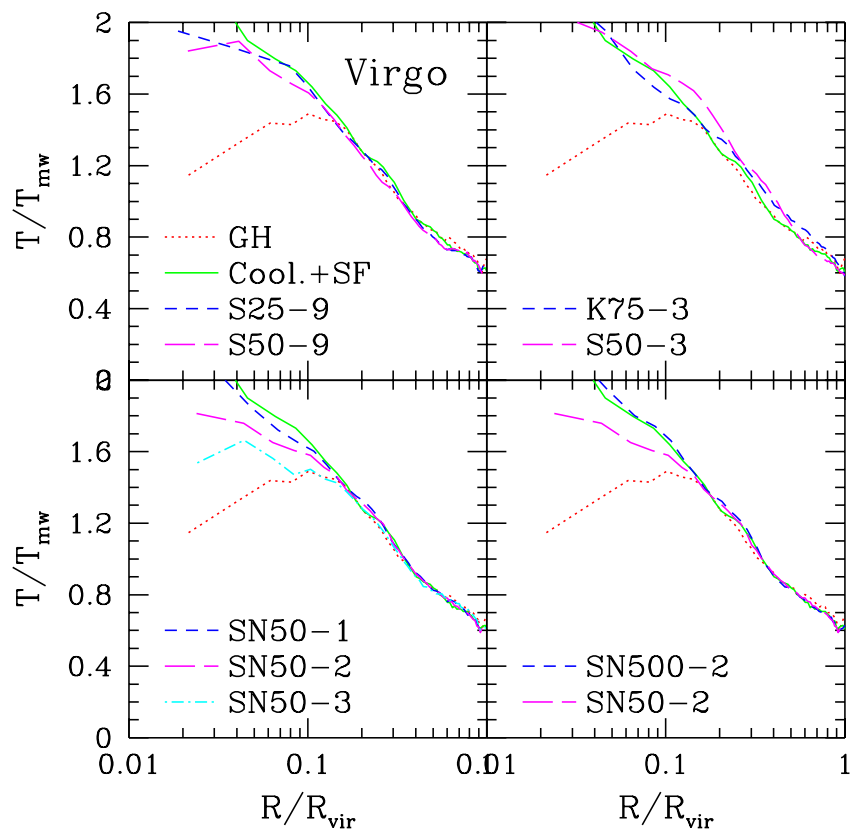,width=12cm} 
}
\vspace{-0.5truecm}
\caption{Temperature profiles of the ``Virgo'' runs, in units of the
mass--weighted temperature. The sequence of panels is the same as in
Figure \ref{fi:sprof}.}
\label{fi:tprof}
\end{figure*}

The way in which cooling acts in reconciling the observed and the
simulated $M$--$T$ relations implies that temperature profiles should
steepen in central cluster regions.  From an observational viewpoint,
the possibility of realizing spatially resolved spectroscopy has
recently opened the possibility to determine temperature profiles for
fairly large samples of clusters. Interestingly, observations based on
the ASCA (e.g., Markevitch et al. 1998) and Beppo--SAX (De Grandi \&
Molendi 2002) satellites show {\em declining} temperature profiles in
the outer regions, at cluster-centric distances $\magcir
0.2$--0.3$R_{\rm vir}$ (cf. also Irwin \& Bregman 2000). This
behaviour is generally reproduced by simulations that do not include
cooling (e.g., BGW). Furthermore, both Beppo--SAX (De Grandi \&
Molendi 2002), Chandra (e.g., Ettori et al. 2002; Allen et al. 2001;
Johnstone et al. 2002) and XMM (e.g., Tamura et al. 2001) data for
fairly hot systems, $T_X\magcir 4$ keV, show temperature profiles
declining towards the very central regions of clusters, thus
indicating the presence of cooling cores.
This behaviour is grossly at variance with respect to that found for
the ``Virgo'' runs, as reported in Figure \ref{fi:tprof}: the only
case where a somewhat declining profile is produced is the one with
gravitational heating, while cooling always gives rise to steeply
increasing profiles with no evidence for any decline, independent of
the presence of extra heating.

A more comprehensive comparison with the observations would require
simulations to be realized for a set of clusters with higher
temperature. On the other hand, our simulated Virgo cluster has been
chosen as a fairly relaxed system. Therefore, as long as observations
suggest profiles to be universal for such systems (Allen et al. 2001),
such a discrepancy should be taken quite seriously. A steepening of
the temperature profiles caused by cooling has been already noticed by
Lewis et al. (2000), Muanwong et al. (2002) and Valdarnini (2002). The
temperature profiles in Fig. \ref{fi:tprof} generalise this result
also in the presence of a variety of extra--heating mechanisms.

We also note that the steep temperature profiles predicted by
simulations are also at variance with respect to those predicted by
the semi--analytical model for ICM heating/cooling by Voit et
al. (2002). A detailed comparison between the predictions of
semi--analytical models and simulations is beyond the scope of this
paper. However, a full understanding of the physical processes taking
place in the ICM will only be obtained if the reasons for such
differences can be understood and eventually sorted out.

If the discrepancy between observed and simulated temperature profiles
will be confirmed, it may indicate that we are missing some basic
physical mechanism which affects the thermal properties of the gas in
the high density cooling regions. For instance, thermal conduction has
been advocated by some authors as a mechanisms that, in combination
with central heating, may regulate gas cooling (e.g. Voigt et al. 
2002) while providing acceptable temperature profiles for a
suitable choice of the conductivity parameter (e.g., Zakamska \&
Narayan 2002; Ruszkowski \& Begelman 2002). In this scenario, one
expects the outer layers to heat gas in the innermost regions, so as
to increase its cooling time, allowing it to stay in the diffuse phase
at a relatively low temperature. However, the detection of sharp
features in the temperature map of several clusters, as observed by
the Chandra satellite, led some authors to suggest that thermal
conduction is suppressed in the ICM (e.g., Ettori \& Fabian
2000). Magnetic fields are naturally expected to produce such a
suppression (e.g., Sarazin 1988). Still, it is not clear whether this
mechanism can act in an ubiquitous way inside clusters or whether the
turbulence associated with the presence of magnetic fields is actually able
to maintain a relatively efficient thermal conduction (e.g., Narayan \&
Medvedev 2001).

\section{Discussion and conclusions}
We presented results from high resolution Tree+SPH simulations of a
moderately poor ``Virgo''--like cluster and of three group--sized
halos, including the effects of radiative cooling and
non--gravitational gas heating. The numerical accuracy reached in
these simulations was aimed at following in detail the pattern of gas
cooling and its effect on the $X$--ray properties of groups and
clusters of galaxies. The main results that we obtained can be
summarised as follows.
\begin{description}
\item[(a)] Including cooling and star formation causes a fraction
  $f_\ast\simeq 0.25$ of baryons to be converted into a collisionless
  ``stellar'' phase in the Virgo cluster and $f_\ast \simeq
  0.35$--0.40 in the simulated groups. Given the sensitivity of
  cooling on numerical resolution, it is likely that the
  result for the ``Virgo'' run should still be interpreted as a lower
  limit on $f_*$.
\item[(b)] The cold fraction is reduced by including
  non--gravitational heating. The degree to which overcooling
  is suppressed depends not only on the amount of feedback
  energy, but also on the redshift and on the gas overdensity at which
  it is released into the diffuse medium. For instance, heating at
  $z_h=9$ is very efficient in decreasing $f_\ast$ below the 10 per
  cent level, at the expense of delaying the bulk of star formation
  to $z\mincir 1$. A more realistic star formation history, peaking at
  $z\simeq 3$, consistently requires that most of the non--gravitational
  heating takes place at a similar redshift, with at least 20 per
  cent of the baryons still being converted into stars.
\item[(c)] Heating at $z_h=3$ with $E_h\simeq 0.75$ keV/part is shown
  to produce scalings of $X$--ray luminosity, mass and entropy
  vs. temperature which agree in general with observational data. This
  result holds independent of whether an equal amount of energy is
  assigned to all gas particles or whether an entropy floor is
  created. A similar agreement is also found for the SAM--predicted SN
  feedback, once realistic models for the IMF are used. Both, heating
  at $z_h=9$, or using an IMF which produces a large number of SN, are
  not efficient in suppressing the $X$--ray luminosity, which is a
  consequence of the fairly large amount of gas that, while avoiding
  cooling, is concentrated in central cluster regions.
\item[(d)] Including cooling and star formation increases the ICM
  temperature in the central regions. While this helps in reconciling
  simulations with the observed $M$--$T$ relation, it steepens
  the temperature profiles, which show no evidence for any decline at
  small cluster-centric distances. This result, which holds independent
  of the scheme for non--gravitational heating, is discrepant
  with recent observations.
\end{description}

Over the last year or so, different groups have presented simulations
aimed at studying the effect of radiative cooling and feedback on the
$X$--ray properties of the ICM. Most of these studies are based on
simulations which follow the gas hydrodynamics within the full volume
of a cosmological box (e.g., Muanwong et al. 2002; Dav\'e et al. 2002;
Kay et al. 2002). One common result of these simulations, which agrees
with what we find in our analysis, is that the effect of cooling is
able to alleviate or even solve the discrepancy between simulated and
observed $X$--ray scaling properties of clusters and groups, but the
fraction of baryons converted into stars is too large. To remedy this
problem, Muanwong et al. (2002) pre--heated the gas by adding 1.5 keV
thermal energy to all the gas particles at $z_h=4$. As a result, they
found that the cold fraction in groups and clusters is decreased from
15 per cent to 0.4 per cent, which is somewhat smaller than the values
found in our simulations. Kay et al. (2002) implemented a feedback
mechanism in their simulations, which accounts for the rate of both
type Ia and type II SN. By assuming an energetics twice as large as
that provided by standard supernova computations, they were able to
reproduce the observed $X$--ray scaling properties, while obtaining
only 3 per cent of the gas to be converted into stars.  The main
limitation of this type of simulations is that one is restricted to
relatively poor numerical resolution in order to limit the
computational cost. For instance, the simulations by Muanwong et
al. (2002) have a mass resolution which is about one order of
magnitude worse than that of our ``Virgo'' runs and almost two orders
of magnitude worse than that of our group runs. A better mass resolution
within a smaller box was used by Kay et al. (2002), for which the mass
of gas particles are a factor 2.8 and 22 smaller than for our Virgo
and Group runs, respectively.

The results that we presented in Section 3 demonstrate that the cooling
efficiency is quite sensitive to mass resolution. For this reason, one
has to be careful in drawing conclusions about overcooling and
how it is suppressed by extra heating, in the presence of limited
numerical resolution. In fact, our simulations demonstrate that the
two main problems caused by the introduction of radiative cooling,
namely the overproduction of stars and the steeply increasing
temperature profiles in central cluster regions, may not be easily
solved by the introduction of non--gravitational heating.

Does this imply that none of our heating schemes is a realistic
approximation to what happens in real clusters?
The energy release in all these schemes misses, although to different
degrees, to faithfully follow the simulated rate of star production. A
realistic scheme for SN feedback should dump thermal energy with a
rate that accurately follows the star formation rate, properly
accounting for the typical life--times of different stellar
populations. Furthermore, our schemes for energy release demonstrate
that for feedback to have a sizeable effect on the ICM thermodynamics,
it has to act in a non--local way, so as to assign most of the energy
on gas particles which have a sufficiently long cooling time. Such
non--local feedback mechanisms may arise from AGN activity, cosmic
rays or galactic winds, for example.

While further work is clearly needed to study such feedback mechanisms
self-consistently in simulations, a better understanding is also
required as to whether optical/$X$--ray data really implies a stellar
fraction as small as $\mincir 10$ per cent within clusters and groups.
Balogh et al. (2001) used the 2MASS results on the K--band luminosity
function by Cole et al. (2001) to estimate the cosmic fraction of
baryons converted into stars. After assuming a Kennicutt IMF
(Kennicutt 1983), they find $f_*\simeq 0.05$ for our choice of
$\Omega_m$ and $h$, and argued that no much evidence exists for
$f_\ast$ to increase inside clusters, or to depend on the cluster mass
(cf. also Bryan 2000). However, this estimate of the cosmic value of
$f_\ast$ increases by about a factor 2 if a Salpeter IMF (Salpeter
1955) were used instead.  Furthermore, it is worth reminding that the
estimate inside clusters relies to some degree of extrapolation. For
instance, Balogh et a. (2001) obtained the stellar mass in clusters
from the B--band luminosity data by Roussel et al. (2000), using
$M/L_B=4.5$, and correcting for undetected galaxies by extrapolating
the luminosity function to the faint end slope. It is clear that a
more robust determination of $f_\ast$ in clusters should rather rely
on K-- or H--band luminosity, which is more directly related to
stellar mass (e.g., Gavazzi et al. 1996), rather than to B--band
luminosity whose conversion to stellar mass is quite sensitive to
galaxy morphology. More recently, Huang et al. (2002) used the
Hawaii-AAO K-band redshift survey to estimate the K-band luminosity
function in the local Universe. They found that the K--band luminosity
density is twice as large as that from 2MASS, thus implying a twice as
large $f_\ast$ value.  In light of this discussion, a $f_*$ value
somewhat larger than 10 per cent, possibly as large as 20 per cent,
may still be viable at present, which would tend to alleviate the
problem of ICM overcooling.

As for the temperature profile, our results indicate that the
discrepancy between observations and simulations is unlikely to be
solved by the inclusion of feedback mechanisms that are similar to the
ones explored here. If this is the case, it would demonstrate that our
simulations are missing some basic physical mechanisms. For instance,
as we discussed, thermal conduction has been proposed to be an
important effect in clusters. Another piece of physics which is
currently missing from most simulation work is the effect of magnetic
fields (e.g., Dolag, Bartelmann \& Lesch 2002). Their introduction
might give rise to non--trivial structures in the gas distribution if
they can locally suppress thermal conduction, or it they provide a
non--thermal contribution to the gas pressure.

There is little doubt that including such more complex physics will
represent a significant, non-trivial challenge for cluster simulations
of the next generation.  Most of the processes involved require both,
a rather sophisticated numerical method, and a treatment of sub-grid
physics. Still, the inclusion of more physics in numerical codes is
mandatory if the reliability and the predictive power of cluster
simulations want to keep pace with the increasing quality of
observational data.

\section*{Acknowledgements}
Simulations were run at the CINECA Supercomputing Center, with CPU
time provided by a grant of the National Institute for Astrophysics
(INAF), at the Computing Center of the Astronomical Observatory of
Catania and at the Computing Center of the University of Trieste. We
wish to thank Hans B\"ohringer, Alexis Finoguenov, Fabio Governato,
Paolo Tozzi and Xian--Ping Wu for enlightening discussions.

\end{document}